\documentclass[11pt,a4paper]{article}
\usepackage[dvipsnames]{xcolor}
\usepackage{jheppub}
\usepackage[T1]{fontenc}
\usepackage[utf8]{inputenc}
\usepackage{lmodern}
\usepackage{enumerate}
\usepackage[english]{babel}
\usepackage{comment}
\usepackage{bm}
\usepackage{float}

\usepackage[normalem]{ulem}

\usepackage{amsfonts}
\usepackage{mathtools}
\usepackage{simplewick}
\usepackage{multirow}
\usepackage{enumitem}
\usepackage{physics}
\usepackage[compat=1.1.0]{tikz-feynman}
\usepackage{subfigure}
\usepackage{booktabs}
\usepackage{tikz}
\usepackage{cancel}
\usetikzlibrary{decorations.pathmorphing}

\allowdisplaybreaks[1]

\definecolor{ccqqqq}{rgb}{1,0.5,0}
\definecolor{uuuuuu}{rgb}{0.26666666666666666,0.26666666666666666,0.26666666666666666}
\definecolor{qqwwzz}{rgb}{0,0.3,0.9}

\newcommand{\beq}{\begin{equation}}
\newcommand{\eeq}{\end{equation}}
\newcommand{\bea}{\begin{eqnarray}}
\newcommand{\eea}{\end{eqnarray}}

\usepackage{adjustbox}
\hypersetup{   colorlinks=true,  citecolor=blue, linkcolor=blue, urlcolor=blue }

\usepackage{changepage}

\title{Entanglement transitions in holographic conformal interfaces}

\author[a,b]{Evangelos Afxonidis,}
\author[c]{Rotem Berman,}
\author[a,b]{Ignacio Carre{\~n}o Bolla,}
\author[c]{Shira Chapman,}
\author[a,b]{Carlos Hoyos,}
\author[c]{and Osher Shoval}

\affiliation[a]{
Departamento de F\'isica,
Universidad de Oviedo,\\
c/ Leopoldo Calvo Sotelo 18, ES-33007, Oviedo, Spain}

\affiliation[b]{
Instituto de Ciencias y Tecnolog\'ias Espaciales de Asturias (ICTEA),\\
c/ Independencia 13, ES-33004, Oviedo, Spain}

\affiliation[c]{Department of Physics, Ben-Gurion University of the Negev,\\
David Ben-Gurion Boulevard 1, Beer Sheva 84105, Israel}

\emailAdd{afxonidisevangelos@uniovi.es}
\emailAdd{bermar@post.bgu.ac.il}
\emailAdd{ignaciocarbolla@gmail.com}
\emailAdd{schapman@bgu.ac.il}
\emailAdd{hoyoscarlos@uniovi.es}
\emailAdd{oshersho@post.bgu.ac.il}

\abstract{We study entanglement transitions for pairs of disjoint intervals in
two-dimensional holographic interface conformal field theories. We consider
several holographic models with equal central charges on the two sides of the
interface: thin-wall geometries with one and two branes, the Janus solution, and
the super-Janus solution. We first discuss the finite interface contribution
to the entanglement entropy of a single interval, distinguishing intervals that
cross the interface from those that lie entirely on one side. We then use the
mutual information to locate the transition between connected and disconnected
Ryu--Takayanagi geodesic configurations for two intervals. When the interface
lies between the two intervals, increasing the boundary entropy $\log g$, which provides a measure
of interface strength in the models considered here, generally
decreases the critical separation at which the connected configuration ceases
to dominate. 
When both intervals lie on the same side of the interface, the
behavior is model dependent: the connected phase persists to larger
separations in the Janus and super-Janus models, while the critical
separation is unchanged in the thin-wall models. Thus, as diagnosed by
the two-interval transition, the interface weakens correlations across
the interface while preserving or enhancing correlations between
regions on the same side. We also
identify a universal ``Silver Blaze'' configuration, in which the interface lies
inside one of the intervals and the transition occurs at the same location as in
a CFT without an interface, independently of the interface parameters and of the
holographic model. This universality follows from a cancellation between
crossing and non-crossing interface contributions to the mutual information. \sloppy}

\begin{document}

\maketitle
\flushbottom

%%%%%%%%%%%%%%%%%%%%%%%%%%%%%%%%%%%%%%%%%%%%%%%%%
%%%%%%%%%%%%%%%%%%%%%%%%%%%%%%%%%%%%%%%%%%%%%%%%%
\section{Introduction}\label{sec:Introduction}
%%%%%%%%%%%%%%%%%%%%%%%%%%%%%%%%%%%%%%%%%%%%%%%%%
%%%%%%%%%%%%%%%%%%%%%%%%%%%%%%%%%%%%%%%%%%%%%%%%%

Entanglement entropy provides a useful probe of how quantum states fail to factorize across spatial subregions of a system.
In conformal field theories
(CFTs), symmetry strongly constrains the entanglement entropy of a single interval. In two dimensions, for example, the vacuum entanglement entropy of a single interval on the infinite line
depends logarithmically on the interval length relative to the UV cutoff, with a coefficient
fixed by the central charge \cite{Calabrese:2004eu,Calabrese:2009qy}.
However, single intervals probe only part of the pattern of quantum correlations in the
state. More refined information is obtained by considering regions made of
multiple disjoint intervals. Already for two disjoint intervals in a two-dimensional CFT, the entanglement
entropy depends on theory-specific CFT data encoded in a four-point function
of twist fields \cite{Caraglio:2008pk,Calabrese:2009ez}. A useful UV-finite quantity in this setting is the mutual information,
which measures the  correlations shared by the two intervals.

\begin{figure}[htbp]
\centering
\subfigure[]{%
  \raisebox{0.25cm}{%
    \includegraphics[scale=0.80]{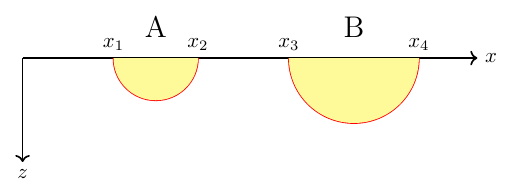}%
  }%
}
\,
\subfigure[]{%
  \includegraphics[scale=0.75]{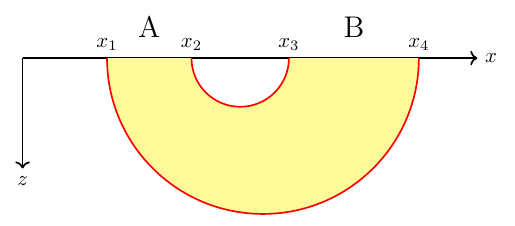}%
}
\caption{Candidate RT geodesic configurations for the boundary region
$\text{A}\cup \text{B}=[x_1,x_2]\cup[x_3,x_4]$ in pure AdS$_3$. 
Panel (a) shows the disconnected configuration, with geodesics homologous to
each interval separately. Panel (b) shows the connected configuration, with
geodesics connecting $x_1$ to $x_4$ and $x_2$ to $x_3$. The dominant
contribution to the entanglement entropy is obtained by choosing the
configuration with smaller total length.}
\label{fig:minimalsurface}
\end{figure}

In holographic CFTs, the Ryu--Takayanagi (RT) prescription relates the entanglement entropy
of a boundary region to the area of an extremal surface in the bulk geometry
\cite{Ryu_2006}. For two disjoint intervals in a two-dimensional holographic CFT, there
are two competing extremal-surface configurations, which in this case are geodesics \cite{Hubeny:2007re}. In the disconnected configuration,
each interval is capped off by its own geodesic. In the connected configuration, the geodesics instead connect endpoints belonging to different intervals. These two configurations are illustrated in Figure \ref{fig:minimalsurface}. The
dominant saddle is the configuration of minimal total length. As the ratio between the
separation and the size of the intervals is varied, the dominant saddle changes
discontinuously, producing a sharp entanglement transition
\cite{Headrick:2010zt}.
Equivalently, the transition can be diagnosed by the mutual information between
the two intervals: it is positive when the connected saddle dominates and
vanishes when the disconnected saddle dominates. Such sharp transitions are
characteristic of holographic CFTs and, in two dimensions, arise in large central charge
CFTs with a sparse spectrum of light operators \cite{Hartman:2013mia}.

The purpose of this work is to understand how this transition is modified in the presence of a conformal interface. Interfaces, defects and boundaries  arise naturally in conformal field theories \cite{Cardy:2004hm,Bachas:2001vj,Billo:2016cpy}.  They describe, among other things, impurities and Kondo-like
systems \cite{Affleck:1995ge}, junctions of quantum wires
\cite{Chamon:2003tz,Oshikawa:2005fh}, and spatial variations of couplings in Janus-type
interfaces \cite{Clark:2004sb}. 
From the point of
view of entanglement, such extended objects are especially interesting because they can
partially obstruct or enhance correlations between different parts of the
system. Boundaries and defects can also carry universal data such as
the boundary entropy in two dimensions, whose behavior under boundary RG flows is constrained by
the \(g\)-theorem \cite{Affleck:1991tk,Friedan:2003yc,Casini:2016fgb}. 
The entanglement entropy of a single interval in the presence of a conformal
interface has been studied in a variety of field-theoretic and holographic settings
\cite{Azeyanagi:2007qj,Sakai:2008tt,Brehm:2015lja,Kruthoff:2021vgv,Karch_2021,
Afxonidis:2024gne,Afxonidis:2025jph,Wang:2026xqp}.

Holography provides a powerful geometric framework for studying strongly coupled CFTs with conformal defects and interfaces. Holographic models of interfaces admit both
bottom-up descriptions and constructions with top-down origins. 
A particularly simple class of bottom-up models is obtained by gluing several
AdS$_3$ regions across one or more constant tension branes. This class includes geometries with one and two branes and is closely related to constructions used in
AdS/BCFT and in brane-world models 
\cite{Karch_2001,Karch_2001_2,Bachas:2001hpy,Takayanagi:2011zk,
Azeyanagi:2007qj,Fujita_2011,Baig:2022cnb}.
Smooth examples with top-down origins are provided by Janus-type solutions, 
where the interface is supported by non-trivial scalar profiles and the
lower-dimensional description can be obtained from higher-dimensional
supergravity constructions. 
This includes non-supersymmetric Janus solutions and half-BPS super-Janus
solutions \cite{Bak_2003,Bak_2007,Chiodaroli_2010}.
The interface data in these models include quantities such as the boundary
entropy \(\log g\)
\cite{Azeyanagi:2007qj,Chiodaroli:2010ur,Karch_2021,Afxonidis:2024gne}
and energy transmission coefficient \({\cal T}\)
\cite{Baig:2022cnb,Bachas:2022etu,Baig:2024hfc}.  
These quantities provide useful
diagnostics of the interface, although they do not in general exhaust its data.
The multiple-interval entanglement transition provides a complementary probe:
one can ask how the presence of the interface changes the competition between
connected and disconnected candidate RT surfaces
and whether in some cases the influence is related to the previous parameters.

In this paper, we study the entanglement transitions for pairs of disjoint
intervals in two-dimensional holographic interface CFTs (ICFTs), with a codimension-one interface
localized at a point on a constant-time slice. We consider several models: the
single-brane thin-wall model, the double-brane thin-wall model, the Janus
solution, and the super-Janus solution. For simplicity, we
restrict to the case in which the CFTs on the two asymptotic sides of the interface have the same central charge.\footnote{Equal central charges on the two sides do not  imply reflection
symmetry of the interface geometry. In particular, the unequal-tension
double-brane thin-wall background is generally asymmetric, although even in this case, the single-interval entropy functions entering our analysis happen to be invariant
under exchanging the two brane tensions.}

Our first step is to determine the finite interface contribution to the entanglement
entropy of a single interval, denoted $\log g^{(2)}$, for both crossing and non-crossing
intervals. The distinction is important: crossing intervals have endpoints on opposite
sides of the interface, while non-crossing intervals lie entirely on one side. In the bulk,
this distinction corresponds to qualitatively different geodesics in the AdS$_2$-sliced
geometry.
We review the derivation of analytic expressions for the single-brane thin-wall model and the super-Janus solution. For the double-brane thin-wall model, we obtain analytic (parametric) formulae that require numerical inversion, while for the Janus solution, we obtain numerical results supplemented by a perturbative expansion.

We then use these results to study the entanglement transition for two intervals. The
transition is located by computing the mutual information associated with the connected
geodesic configuration and finding where it vanishes. When the interface lies between the
two intervals, we find that 
increasing $\log g$, which provides a measure
of interface strength in the models considered here, generally decreases the
critical separation at which the transition occurs. Equivalently, the system enters the
disconnected phase when the two intervals are closer together than in the CFT without an interface. From the bulk point of
view, the interface effectively stretches space and thus increases the cost of geodesics that connect the two sides,
thereby disfavoring the connected saddle. In the large-$\log g$ limit, the critical separation tends to zero.

The behavior can be reversed when the two intervals lie on the same side of the interface. In the Janus and super-Janus models,
increasing the boundary entropy $\log g$, increases the critical separation, so that the connected phase persists over a larger range. In the 
thin-wall models, by contrast, the relevant geodesics remain entirely within the exterior AdS region and the critical separation is unchanged. Thus, as diagnosed by the two-interval transition, the interface does not simply suppress correlations uniformly: correlations across the interface are weakened, while correlations between regions on the same side are either preserved or enhanced. The detailed dependence on $\log g$ and on the additional interface parameters is model-dependent.

A particularly simple and universal result emerges when the interface lies inside one of
the intervals. 
Each of the two competing configurations then contains one crossing
and one non-crossing contribution. In this case, there exists a special configuration of the four endpoints for
which the transition occurs at precisely the same location as in a CFT without an
interface, independently of the interface parameters, and independently of the holographic
model. If the interface is placed at the origin and the endpoints are ordered as
$x_1<0<x_2<x_3<x_4$, this configuration is characterized by
\begin{equation}
        -x_1 = x_3, \qquad x_3=\sqrt{x_2 x_4}.
\end{equation}
At this point, the crossing and non-crossing interface contributions to the mutual
information cancel pairwise. We refer to this arrangement as the {\em Silver Blaze}
configuration.\footnote{We are borrowing the name from the Silver Blaze problem of QCD with a chemical potential \cite{Cohen:2003kd,Cohen:2026pzh}. {\em Silver Blaze} \cite{Doyle1894} is a Sherlock Holmes story where the essential clue is that a dog did nothing during the night. Similarly, in the Silver Blaze configuration it is as if the interface is not doing anything. In addition, the critical separation of equal length intervals is determined by the silver ratio. So the name fits even better than in QCD. 
} Its existence follows only from the structure of the competing geodesic
configurations and the scale symmetry of the conformal interface, and is therefore
universal within the class of holographic ICFTs considered here.

The rest of the paper is organized as follows. In Section~\ref{sec:HolographicICFTs},
we review the holographic ICFT models used throughout the paper: the one-brane and
two-brane thin-wall geometries, the Janus solution, and the super-Janus solution.
In Section~\ref{sec:HolographicEE}, we compute the entanglement entropy of single
intervals in these backgrounds and extract the finite interface contribution
$\log g^{(2)}$ for crossing and non-crossing intervals. In
Section~\ref{sec:EntanglementTransition}, we study the entanglement transition for two
intervals, determine the critical curves in the different models, and identify the
universal Silver Blaze configuration. We conclude in Section~\ref{sec:Discussion} with
a discussion of the results and several directions for future work. 
The appendices contain the technical details of the geodesic computation
(appendix~\ref{sec::appendixGeodComp}) and an alternative geometric derivation
for thin-wall models (appendix~\ref{sec:AppendixGeometricThinBrane}).

\section{Holographic ICFTs}\label{sec:HolographicICFTs}
%%%%%%%%%%%%%%%%%%%%%%%%%%%%%%%%%%%%%%%%%%%%%%%%%
%%%%%%%%%%%%%%%%%%%%%%%%%%%%%%%%%%%%%%%%%%%%%%%%%

In this section, we describe general properties of holographic duals of
$1+1$ dimensional ICFTs and review the explicit examples used throughout
the paper. We focus on thin-wall 
models with one brane
\cite{Karch_2001,Karch_2001_2,Bachas:2001hpy,Azeyanagi:2007qj}
and two branes \cite{Baig:2022cnb}, the Janus solution
\cite{Bak_2003,Bak_2007}, and the super-Janus solution
\cite{Chiodaroli_2010}.

Holographic duals of $1+1$ dimensional ICFTs are commonly described by
three-dimensional geometries. The thin-wall examples are bottom-up
constructions, in which locally AdS\(_3\) regions are glued across one or
more constant-tension branes. By contrast, the Janus-type examples have
top-down origins. They arise from higher-dimensional supergravity solutions whose
reduction over a compact internal manifold gives an effective lower-dimensional description with non-trivial scalar
profiles. The examples relevant below include non-supersymmetric Janus
solutions and half-BPS super-Janus solutions
\cite{Bak_2003,Bak_2007,Chiodaroli_2010}.

In all the examples considered below, the effective three-dimensional geometry
can be written conveniently in an AdS$_2$ slicing,
\begin{equation}\label{eq::defectmetric}
ds^2=R^2 \left(dr^2+e^{2A(r)}\frac{dx^2-dt^2}{x^2}\right) \ .
\end{equation}
Here $r$ labels the AdS$_2$ slices, while $(t,x)$ are coordinates on each
slice. The boundary of a given AdS$_2$ slice is located at $x=0$. For the special choice $e^{A(r)}=\cosh r$, the metric describes a pure AdS$_3$ spacetime with radius $R$. 
Indeed, performing the coordinate transformation
\begin{equation}\label{eq:poincarecoord}
z=\frac{x}{\cosh r}\ , \qquad y=x\tanh r \ 
\end{equation}
brings the metric to the standard Poincar\'e form
\begin{equation}\label{eq::Poincarepatch}
ds^2=\frac{R^2}{z^2}\left(dz^2-dt^2+dy^2\right) \ .
\end{equation}
In these coordinates, the asymptotic boundary is at $z=0$, and $y$ is the
spatial coordinate of the boundary CFT. The two limits $r\to+\infty$ and
$r\to-\infty$, with $x>0$ fixed, approach the two half-planes $y>0$ and
$y<0$, respectively. The limit $x\to0$, instead, reaches the line
$y=0$.

For a generic warp factor, the two asymptotic
regions \(r\to\pm\infty\) with $x>0$ fixed correspond to the two sides of the interface, while
the interface itself is localized at $x=0$. The warp factor typically has a minimum value 
$e^{A_*}$ at
some position $r=r_*$. This minimum will play an important role in the
geodesic analysis of Section~\ref{sec:HolographicEE}.
We refer to the point $r_*$ as the turnaround point.

%%%%%%%%%%%%%%%%%%%%%%%%%%%%%%%%%%%%%%%%%%%
\subsection{Thin wall model with one brane}\label{sec:Thinwall1}
%%%%%%%%%%%%%%%%%%%%%%%%%%%%%%%%%%%%%%%%%%%

A particularly instructive and widely studied example is the thin wall model with one brane \cite{Randall_1999_1,Randall_1999_2,Karch_2001,Karch_2001_2}. See for instance \cite{Azeyanagi:2007qj,Karch_2021,Afxonidis:2024gne} for analyses of the interface entropy. This setup consists of Einstein gravity with a negative cosmological constant coupled to a matter source localized on a brane. 
The brane is modeled as a thin codimension-one hypersurface  
carrying constant energy density. 
The relevant action reads:
\begin{equation}\label{eq:thinbranesymaction}
    S=\frac{1}{16\pi G} \int d^3x\,\sqrt{-g}\left(\mathcal{R}+\frac{2}{{R}^{2}}\right)-\frac{\Sigma}{8\pi \, G}\int d^2x\,\sqrt{-h} \ ,
\end{equation}
where $\mathcal{R}$ is the Ricci scalar, $R$ is the AdS radius, $\Sigma>0$ is the tension, $G$ is Newton's constant and $h$ is the induced metric on the brane. 
When the tension lies below a critical value, $\Sigma\leq 2/R$, the brane intersects the asymptotic AdS boundary. The bulk geometry is then composed of two pure AdS regions glued together at the brane location. Such a configuration naturally provides a bottom-up holographic realization of an ICFT, although without a known explicit embedding into string theory. Similar constructions also arise in the description of higher-dimensional impurities after performing an $s$-wave reduction \cite{Erdmenger:2013dpa,Erdmenger:2015spo,Erdmenger:2015xpq}.

Within our framework, the thin wall geometry corresponds to the warp factor
\begin{equation}\label{eq::warpfactorTW}
e^{A}=\cosh{\left(|r|-r_*\right)} \ , \qquad r_* = \tanh^{-1}\left(\frac{\Sigma R}{2}\right) \ ,
\end{equation}
which is defined piecewise for positive and negative values of $r$. The derivative is discontinuous at $r=0$, signaling the presence of the brane at that location. Unlike the pure AdS case, the minimum value of the warp factor, $e^{A_*}=1$, is reached at $r=\pm r_*$ rather than at $r=0$. 

The boundary entropy for the thin-brane geometry is given by \cite{Takayanagi:2011zk}
\begin{equation}\label{eq::boundaryentropyRS}
\log g=\frac{R r_*}{2G}=\frac{cr_*}{3}=\frac{c}{3}\tanh^{-1}{\left(\frac{\Sigma R}{2}\right)} \ ,
\end{equation}
where $c$ denotes the central charge of the CFT, and we have used the Brown-Henneaux formula \cite{Brown:1986nw}
\begin{equation}\label{eq:centralc}
c=\frac{3R}{2G}\ .
\end{equation}

%%%%%%%%%%%%%%%%%%%%%%%%%%%%%%%%%%%%%%%%%%
\subsection{Thin wall model with two branes}\label{sec:Thinwall2}
%%%%%%%%%%%%%%%%%%%%%%%%%%%%%%%%%%%%%%%%%%%

We now consider a generalization of the thin wall setup in which the bulk spacetime is divided into three locally AdS$_3$ wedges separated by two codimension one branes. We denote these regions by $i=L,C,R$ corresponding respectively to the left, central and right wedge. The AdS radius could take different values in each of these three regions, but we restrict our analysis to configurations in which all of them  have the same AdS radius. 
The metric can be written as
\begin{equation}\label{eq::defectmetricDTW}
ds^2=R^2 \left(dr^2+e^{2A_i(r)}\frac{dx^2-dt^2}{x^2}\right) \ ,
\end{equation}
where $R$ is the common AdS radius and $A_i(r)$ is the warp factor
in region $i$.

We assume the metric in each of the regions is a portion of empty AdS. 
Without loss of generality, we can always shift $r$ such that the warp factor takes the following piecewise form:
\begin{equation}\label{eq::warpfactortwoTW}
 \, e^{A_i(r)}=\left\{ \begin{array}{lclc}
    \cosh{(r-a)} \, &, \quad & r_1>r>-\infty \, & \quad \ (i=L)\\
    \cosh{r} \, &, \quad & r_2>r>r_1\, & \quad \ (i=C) \\
    \cosh{(r-b)}\,   &, \quad & +\infty>r>r_2\, & \quad \ (i=R)
\end{array}\right. \ .
\end{equation}

The parameters $r_1,r_2,a,b$ are fixed by the continuity of the induced metric at the location of the branes together with the Israel junction conditions \cite{Israel:1966rt,Baig:2022cnb}:
\begin{subequations}\label{eq::IsraelJunction}
\begin{align}
       \cosh{(r_1-a)}&= \cosh{r_1},\quad  \cosh{(r_2-b)}= \cosh{r_2},\\
       -\frac{1}{R} \tanh{(r_1-a)}+  \frac{1}{R} \tanh{r_1}&= -\Sigma_1,\quad
      \frac{1}{R} \tanh{(r_2-b)}-  \frac{1}{R} \tanh{r_2}= -\Sigma_2.
\end{align}
  \end{subequations}
 Here $\Sigma_1$ and $\Sigma_2$ are the tensions of the branes located at $r_1,r_2$. 
A finite real solution exists provided \cite{Baig:2022cnb}
\begin{equation}\label{eq:existence_cond}
0<\Sigma_i R<2\,,
\qquad i=1,2 .
\end{equation}
Solving the Israel junction conditions determines the shifts $a,b$ and the brane positions $r_1,r_2$ in terms of the brane tensions and the AdS radius:
\begin{equation}\label{eq::solIsraelJunction}
    r_1=- \tanh^{-1}\left(\frac{ R\Sigma_1}{2}\right)\,,\quad
        r_2= \tanh^{-1}\left(\frac{ R\Sigma_2}{2}\right)\,, \quad a=2r_1\,,\quad b=2 r_2 \ .
\end{equation}
Using \eqref{eq::solIsraelJunction}, one has
\(a<r_1<0<r_2<b\), so each branch of the piecewise warp factor
\eqref{eq::warpfactortwoTW} contains the minimum of its corresponding
\(\cosh\). The full warp factor therefore has three degenerate global
minima, at \(r=a,0,b\), all with \(e^{A_*}=1\).

The double thin-wall model considered here has two independent
dimensionless parameters, which may be taken to be \(R\Sigma_1\) and
\(R\Sigma_2\), in addition to the central charge. 
It is convenient to replace the two tensions by their sum and difference,
$R(\Sigma_2+\Sigma_1)$ and $R(\Sigma_2-\Sigma_1)$. 
The sum of the tensions has a direct interpretation in terms of the energy
transmission coefficient, which in the double thin-wall model is given
by~\cite{Baig:2022cnb,Bachas:2022etu}
\begin{equation}
    \mathcal{T}
    =
    \frac{1}{1+\frac{R}{2}(\Sigma_2+\Sigma_1)}.
\end{equation}
Another useful physical parameter for the double thin-wall model is the boundary entropy given by \cite{Baig:2022cnb} 
\begin{equation} \label{eq::logg2TW}
      \log g=\frac{c}{6}\left(b-a\right)\,.
\end{equation}
In terms of the tensions this is
\begin{equation}
    \log g=\frac{c}{3}\left(\tanh^{-1}\left(\frac{ R\Sigma_1}{2}\right)+\tanh^{-1}\left(\frac{ R\Sigma_2}{2}\right)\right)\,.
\end{equation}
Thus, $\log g$ 
and $\mathcal{T}$ provide two physical observables associated
with the double thin-wall interface 
considered here. We use $\log g$ together with $R(\Sigma_2-\Sigma_1)$ as the independent parameters in our plots later on. More general double thin-wall constructions can involve additional
independent parameters, for example a distinct AdS radius in the central
wedge. We leave the study of such configurations for future work.

When the tensions are the same $R(\Sigma_2-\Sigma_1)=0$, the value of $\log g$ can be made arbitrarily small by taking small values of the tensions. However, when the tensions are different, for a fixed value of the difference $R(\Sigma_2-\Sigma_1)\neq 0$, the boundary entropy has a minimal value that is reached when one of the tensions is taken to zero, but both cannot be made zero simultaneously. This implies a lower bound
\begin{equation}\label{eq::boundaryentrbound}
    \log g \geq \frac{c}{3}\tanh^{-1}\left|\frac{R(\Sigma_2-\Sigma_1)}{2}\right| \ .
\end{equation}
This bound will be relevant when we study the entanglement transition, since we will study the location of the transition for fixed values of $R(\Sigma_2-\Sigma_1)$ as a function of $\log g$.

%%%%%%%%%%%%%%%%%%%%%%%%%%%%%%%%%%%%%%%%%%%
\subsection{Janus solution}\label{sec:Janus}
%%%%%%%%%%%%%%%%%%%%%%%%%%%%%%%%%%%%%%%%%%%

We turn next to the Janus solution, originally introduced as a supergravity dual of a $3+1$-dimensional ICFT \cite{Bak_2003}.  
A realization of the dual of a $1+1$ dimensional ICFT was later constructed in \cite{Bak_2007}.  
Upon reduction of ten-dimensional supergravity to three dimensions, the dual to the $1+1$ ICFT is a solution of Einstein gravity coupled to a dilaton field $\phi$  with a flat potential corresponding to the action \cite{Chiodaroli_2010}: 
\begin{equation}
S=\frac{1}{16\pi G}\int d^{3}x\,\sqrt{-g}\left[\mathcal{R}-\partial^\mu\phi\,\partial_\mu\phi+\frac{2}{R^2}\right].
\end{equation}
The metric takes the form \eqref{eq::defectmetric} with warp factor  
\begin{equation}
e^{2A(r)}=\frac{1}{2}\left(1+\left( 1-\beta\right)\cosh{(2r)} \right) \ ,
\end{equation} 
where the parameter satisfies $0\leq \beta\leq 1$. The dilaton profile is given by 
\begin{equation}
\phi(r)=\phi_0+\frac{1}{\sqrt{2}}\log{\left( \frac{\sqrt{2-\beta}+\sqrt{\beta}\tanh{r}}{\sqrt{2-\beta}-\sqrt{\beta}\tanh{r}}\right)} \ .
\end{equation}
The parameters $\phi_0$ and $\beta$ dictate the values $\phi(\pm \infty)$ which fix the marginal couplings of the CFT on the two sides of the interface. For $\beta=0$, the dilaton becomes constant and proportional to $\phi_0$, while the geometry reduces to pure AdS$_3$ without an interface.  In the limit $\beta\rightarrow 1$, the spacetime degenerates into $\mathbb{R}\times$AdS$_2$. The minimum value of the warp factor in the Janus background is
\begin{equation}\label{eq::minimalwarpJanus}
e^{A_*}=\sqrt{1-\frac{\beta}{2}} \ .
\end{equation}

The boundary entropy is \cite{Azeyanagi:2007qj}\footnote{In order to compare with the usual parametrization of the Janus solution one should take $\beta=1-\sqrt{1-2\gamma^2}$.}
\begin{equation}\label{eq:boundentropyJ}
    \log g=-\frac{R}{4G}\log(1-\beta)=-\frac{c}{6}\log(1-\beta) \ .
\end{equation}

%%%%%%%%%%%%%%%%%%%%%%%%%%%%%%%%%%%%%%%%%%%
\subsection{Super Janus solution}\label{sec:SuperJanus}
%%%%%%%%%%%%%%%%%%%%%%%%%%%%%%%%%%%%%%%%%%%

A Janus solution dual to a $1+1$ dimensional supersymmetric ICFT was first constructed in \cite{Chiodaroli_2010}. 
The boundary entropy of this solution was computed in \cite{Chiodaroli:2010ur}. More recently, \cite{Baig:2024hfc} performed a KK reduction to an effective three-dimensional description and used it to compute the transmission coefficient. The ten-dimensional solution contains a dimensionful parameter $L$ fixing the central charge and two dimensionless  parameters $\psi$,$\theta$, which parametrize the jump of the dilaton and axion  across the interface.   
In the reduced description these parameters determine the effective AdS$_3$ radius and warp factor, yielding a metric of the form \eqref{eq::defectmetric} with 
\begin{equation}\label{eq:wrapSJ}
R^2=2L\cosh{\psi}\cosh{\theta}\ , \quad\quad e^{A(r)}=\frac{\cosh{r}}{\cosh{\psi}\cosh{\theta}} \ .
\end{equation} 
The minimum of the warp factor in the super Janus geometry occurs at $r=0$ and is given by
\begin{equation}
e^{A_*} = \frac{1}{\cosh \psi \cosh \theta}= \sqrt{\mathcal{T}}.
\end{equation}  
The boundary entropy is given by \cite{Chiodaroli:2010ur}
\begin{equation}\label{eq:boundarygSuperJanus}
    \log g=-\frac{R}{4G}\log \mathcal{T}=-\frac{c}{6}\log\mathcal{T} \ .
\end{equation}
Note that the warp factor is proportional to that of AdS$_3$ and coincides with it for $\mathcal{T}=1$. One can use this fact to make the super Janus metric coincide with that of AdS$_3$ with radius $R$ \emph{on a constant time slice}. Indeed, with the change of coordinates
\begin{equation}\label{eq:changecoordSJ}
    u=\operatorname{sign}(x)|x|^{\sqrt{\mathcal{T}}},
\end{equation}
the metric becomes Lifshitz-like with dynamical exponent 
$\mathcal{Z}=1/\sqrt{\mathcal{T}}$:
\begin{equation}
    ds_{\rm sJ}^2=R^2\left(dr^2+\cosh^2r\left(\frac{du^2}{u^2}-\mathcal{T}\frac{dt^2}{u^{2\mathcal{Z}}}\right)\right).
\end{equation}

%%%%%%%%%%%%%%%%%%%%%%%%%%%%%%%%%%%%%%%%%%%%%%%%
%%%%%%%%%%%%%%%%%%%%%%%%%%%%%%%%%%%%%%%%%%%%%%%%
\section{Entanglement entropy in holographic ICFTs}\label{sec:HolographicEE}
%%%%%%%%%%%%%%%%%%%%%%%%%%%%%%%%%%%%%%%%%%%%%%%%
%%%%%%%%%%%%%%%%%%%%%%%%%%%%%%%%%%%%%%%%%%%%%%%%

The entanglement entropy (EE) associated with a spatial region A is
$S_{\text{A}}=-\mathrm{Tr}\,\rho_\text{A}\log\rho_\text{A}$, where
$\rho_\text{A}=\mathrm{Tr}_{\bar{\text{A}}}\rho$ is the reduced density matrix. In
holographic theories, it can be computed using the Ryu-Takayanagi (RT)
prescription \cite{Ryu_2006}, which states that
\begin{equation}
S_\text{A}=\frac{\mathrm{Area}(\gamma_\text{A})}{4G},
\end{equation} where $\gamma_\text{A}$ is the codimension-two
extremal surface anchored on the entangling surface $\partial \text{A}$ at the asymptotic AdS boundary and
homologous to A. For a $1+1$-dimensional field theory, and on a
constant-time slice of the dual geometry, this surface is simply a spacelike
geodesic connecting the endpoints of the boundary interval A. We review
here the calculation in ICFTs, following closely \cite{Afxonidis:2025jph}.

%%%%%%%%%%%%%%%%%%%%%%%%%%%%%%%%%%%%%%%%%%%%%%%%
\subsection{Minimal length geodesics and interface entropy}\label{sec:GeodConfInterEntr}
%%%%%%%%%%%%%%%%%%%%%%%%%%%%%%%%%%%%%%%%%%%%%%%%

We now summarize the construction of minimal geodesics in the presence of an interface, following the analysis of \cite{Karch_2021,Afxonidis:2024gne,Afxonidis:2025jph}. We will focus on the case of equal central charges for the CFTs residing at each side of the interface. Technical details of the derivation of the geodesic profile and the cutoff prescription are collected in Appendix \ref{sec::appendixGeodComp}.

An interval is characterized by the position of its endpoints relative to the interface. We denote by $l_L$ and $l_R$ the distances from the interface to the left and the right endpoints respectively. We refer to intervals whose endpoints lie on opposite sides of the interface as \emph{crossing} intervals, and to intervals whose endpoints lie on the same side as \emph{non-crossing} intervals, according to whether the associated bulk geodesics connect the two sides of the interface.

With the dual ICFT metric introduced in \eqref{eq::defectmetric}, a geodesic on a constant time slice may be described by a profile $x(r)$. The induced metric on the slice is invariant under a constant rescaling $x\to \lambda x$. This symmetry implies the existence of a conserved Noether charge $c_s$. 
The profile satisfies the first order equation 
\begin{equation}\label{eq::diffeq}
    \frac{x'}{x}=\pm \frac{|c_s| e^{-A}}{\sqrt{ R^2 \,e^{2A}-c_s^2}} \ .
\end{equation}
We take $c_s \geq 0$ without loss of generality and use the sign choice in \eqref{eq::diffeq} to distinguish the two possible branches of the geodesic solution. In particular, as can be seen from equations \eqref{eq::logratioequation}-\eqref{eq::logratioequation2} below, the positive branch of \eqref{eq::diffeq} corresponds to $l_R\geq l_L$, while the negative branch corresponds to $l_L\geq l_R$. The qualitative behavior of the solution is controlled by the value of $c_s$ relative to the minimum warp factor, $e^{A_*}$:
\begin{itemize}
\item If $c_s=0$, one finds $x'=0$, implying $l_L=l_R=l/2$, corresponding to crossing intervals that are symmetric with respect to the interface. 
\item If $c_s < R\, e^{A_*}$, the denominator never vanishes and the solution has no radial turning point. Such geodesics connect the two asymptotic regions and describe crossing intervals. 
\item If instead $c_s > R\, e^{A_*}$, the geodesic reaches a turning point at $r=r_{\text{turn}}$  determined by $c_s=R\, e^{A(r_{\text{turn}})}$ and the full solution is obtained by gluing the two branches at $r=r_{\text{turn}}$. Such geodesics describe non-crossing intervals. 
\item The limiting case $c_s=R\,e^{A_*}$ corresponds to an interval with one endpoint located at the interface. 
\end{itemize}

\begin{figure}[htbp] 
    \centering
    \includegraphics[width=0.95\textwidth]{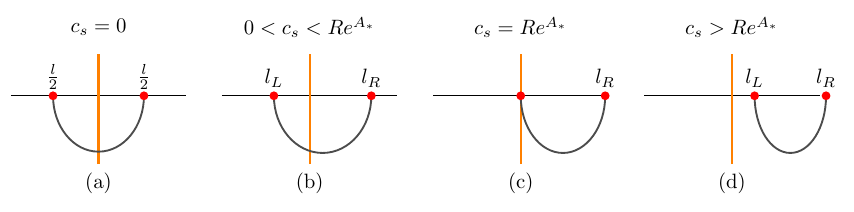}
    \caption{We qualitatively depict here the four cases described in the main text, where the orange vertical line is the interface at $r=0$ and the horizontal black line represents the boundary where the CFT resides. We depict the fully symmetric case $c_s=0$ (a), the generic interface crossing case $0<c_s< R\, e^{A_*}$ (b), the interface touching case $c_s= R\, e^{A_*}$ (c) and finally the non-crossing case $c_s>R\,  e^{A_*}$ (d).}
    \label{fig:interface_cases}
\end{figure}

All possible minimal geodesic configurations are shown schematically in Figure \ref{fig:interface_cases}.

For crossing intervals, the scaling charge $c_s$ is fixed implicitly by the ratio of the endpoint distances to the interface as 
\begin{equation}\label{eq::logratioequation}
    \log\left( \frac{l_L}{l_R}\right)=\mp\int_{-\infty}^{\infty}dr\,\frac{c_s e^{-A}}{\sqrt{e^{2A}R^2-c_s^2}}\ , 
\end{equation}
where the sign choice specifies the branch selected in \eqref{eq::diffeq}. For geodesics with a turning point at $r_{\text{turn}}$, the analogous relation becomes 
\begin{equation}\label{eq::logratioequation2}
    \log\left( \frac{l_L}{l_R}\right)=-2\int_{r_{\text{turn}}}^\infty\frac{c_s e^{-A}}{\sqrt{e^{2A}R^2-c_s^2}}dr \ , \quad \log\left( \frac{l_L}{l_R}\right)=+2\int_{-\infty}^{r_{\text{turn}}}\frac{c_s e^{-A}}{\sqrt{e^{2A}R^2-c_s^2}}dr \ ,
\end{equation}
where each expression corresponds to whether the non-crossing interval is to the right of the interface $l_L\leq l_R$ or to the left $l_L\geq l_R$. In reflection symmetric geometries, the relation \eqref{eq::logratioequation2} may also be used for crossing intervals by setting $r_{\text{turn}}=0$.

Next, we state the corresponding entanglement entropies for crossing and non-crossing intervals. We implement the standard  
regularization prescription. With our conventions for the geometry, the asymptotic expansion of the warp factor is
\begin{equation}
    e^{A}\underset{r\to \pm \infty} \sim \frac{1}{2}e^{a_\pm}e^{\pm r}\,.
\end{equation}
In pure AdS $a_+=a_-=0$.  We introduce a cutoff in the radial direction that corresponds to a UV cutoff $\epsilon$ in the dual field theory (see Appendix \ref{sec::appendixGeodComp} for details). 

The entanglement entropy can be written as
\begin{equation}\label{eq:SA}
    S_{\rm A}=\frac{c}{6}\log \frac{2l_L}{\epsilon}+\frac{c}{6}\log \frac{2l_R}{\epsilon}+\log g^{(2)} \ ,
\end{equation}
 We refer to $\log g^{(2)}$ as the \emph{finite contribution} to the entanglement entropy from now on. The finite contribution for a symmetric interval centered on the interface is $\log g^{(2)}=\log g$, which is the quantity analogous to the boundary entropy in a BCFT, and is given by \cite{Afxonidis:2025jph}
\begin{equation}\label{eq::loggapam}
    \log g=-\frac{c}{6}(a_++a_-)\;.
\end{equation}

For a generic crossing interval, the finite interface contribution is 
\begin{equation} \label{eq::loggicc}
    \log g^{(2)}_c =\frac{c}{6}\int_{-\infty}^{\infty}dr\left(\frac{1}{\sqrt{1-c_s^2e^{-2A}R^{-2}}}-1 \right)+\log g\ .
\end{equation}
For a non-crossing interval on the right of the interface, the finite interface contribution is given by\footnote{The formula for a non-crossing interval on the left of the interface can be found in \eqref{eq::loggicc2app1}.}
\begin{equation} \label{eq::loggicc2}
    \log g^{(2)}_{nc} =\frac{c}{3}\left[\int_{r_{\text{turn}}}^{\infty}dr\left(\frac{1}{\sqrt{1-c_s^2e^{-2A}R^{-2}}}-1 \right)-r_{\text{turn}}\right]-\frac{ca_+}{3}\ .
\end{equation}
These expressions provide the quantities that will be evaluated in the examples below.

Note that all the formulas above assume that the AdS radius is the same everywhere and that the warp factor is expressed as a continuous function of a single coordinate $r$ (rather than in patches).

%%%%%%%%%%%%%%%%%%%%%%%%%%%%%%%%%%%%%%%%%%%%%%%%
\subsection{Examples}\label{sec:formulasModels}
%%%%%%%%%%%%%%%%%%%%%%%%%%%%%%%%%%%%%%%%%%%%%%%%

We collect here the results for the finite contribution to the entanglement entropy in the Holographic ICFTs that were listed in Section \ref{sec:HolographicICFTs}. We will use the dimensionless ratio
\begin{equation}\label{eq:ratiolllr}
    \rho\equiv l_L/l_R
\end{equation}
in the formulas to make them look less cluttered. It should be noted that with the parametrization given in \eqref{eq:SA}, the finite part is non-zero even in a CFT with no interface. We will start reviewing this case, i.e., the case where the geometry is AdS.

%%%%%%%%%%%%%%%%%%%%%%%%%%%%%%%%%%%%%%%%%%%%%%%%
\subsubsection{AdS geometry}
%%%%%%%%%%%%%%%%%%%%%%%%%%%%%%%%%%%%%%%%%%%%%%%%

As a sanity check, we can employ the formalism discussed above and compute the finite  contributions to the entanglement entropy when there is no interface. In that case, the metric \eqref{eq::defectmetric} should be AdS$_3$. As mentioned before, this corresponds to the warp factor
 $   e^{A}=\cosh{r}$. 
Then, one can integrate \eqref{eq::logratioequation} or \eqref{eq::logratioequation2} and solve for $c_s$ in terms of $\rho$ for crossing and non-crossing intervals 
\begin{equation}
\frac{c_s}{R}=\frac{1-\rho}{1+\rho}
\quad \text{(crossing)},\qquad
\frac{c_s}{R}=\frac{1+\rho}{1-\rho}
\quad \text{(non-crossing)}.
\end{equation}  
Substituting the result into the corresponding finite interface contribution  \eqref{eq::loggicc} or \eqref{eq::loggicc2}, yields 
\begin{equation}\label{eq::pureAdS}
   \log g_{c}^{(2)}=-\frac{c}{6}\log\frac{4 \rho}{(1+\rho)^2}\,, \quad 
   \log g_{nc}^{(2)}=-\frac{c}{6}\log\frac{4 \rho}{(1-\rho)^2}\,.
\end{equation}
Even though it looks like there is a non-trivial dependence on the ratio $\rho$ in the finite contribution, when this is combined with the divergent contribution in \eqref{eq:SA}, one recovers the usual CFT result for the entanglement entropy
\begin{equation}
    S_\text{A}=\frac{c}{3}\log \frac{l}{\epsilon} \ ,
\end{equation}
where $l$ is the total length of the segment.

%%%%%%%%%%%%%%%%%%%%%%%%%%%%%%%%%%%%%%%%%%%%%%%%
\subsubsection{Thin wall with one brane}
%%%%%%%%%%%%%%%%%%%%%%%%%%%%%%%%%%%%%%%%%%%%%%%%

We can compute the finite interface  contribution $\log g^{(2)}$ for the thin-wall model with one brane following similar steps as in the AdS geometry. The warp factor in this case is given in \eqref{eq::warpfactorTW}. The result is
\begin{subequations}\label{eq::logg2RS}
\begin{align} 
\label{eq::logg2RSc}    \log g_c^{(2)}&=-\frac{c}{6}\log\frac{4 \rho}{((1+\rho)\cosh r_*+ 2 \sqrt{\rho} \sinh r_*)^2},\;\\
\label{eq::logg2RSnc}    \log g_{nc}^{(2)}&=-\frac{c}{6}\log\frac{4 \rho}{(1-\rho)^2}\,.
\end{align}
\end{subequations}
Notice that for a non-crossing configuration \(\log g^{(2)}\) becomes
independent of \(r_*\) and agrees with the pure-AdS result
\eqref{eq::pureAdS}. For positive tension and equal AdS radii, each
side of the thin wall contains more than half of a copy of AdS, so a
geodesic anchored to a non-crossing interval cannot cross the brane and
return. The geodesic therefore remains entirely within a single AdS
patch. Since the metric in that patch can be brought to the pure-AdS
form by a coordinate transformation, all dependence on the interface
drops out. For the Janus and super-Janus geometries discussed below, the
non-crossing contribution can instead depend nontrivially on the
interface parameters. Within the thin-wall model, relaxing the
equal-radius assumption may permit same-side brane-crossing geodesics,
as in \cite{Anous_2022}. We leave the study of this more general setup
for future work.

An alternative geometric picture for the geodesics in the thin-wall model is presented in Appendix \ref{sec:AppendixGeometricThinBraneSingle}, which provides some useful intuition.

%%%%%%%%%%%%%%%%%%%%%%%%%%%%%%%%%%%%%%%%%%%%%%%%
\subsubsection{Thin wall with two branes}\label{sec:twoTW}
%%%%%%%%%%%%%%%%%%%%%%%%%%%%%%%%%%%%%%%%%%%%%%%%

As mentioned before, we restrict the analysis to the case of equal AdS radii for the three AdS wedges. 
As discussed in \cite{Afxonidis:2025jph}, we need to impose smoothness of the geodesic profile \eqref{eq::diffeq} across each brane:
\begin{subequations}\label{eq::smoothcond}
    \begin{align}
    \label{eq::smoothcond1}\frac{x'_L}{x_L}\Bigg|_{r_1}&=\frac{x'_C}{x_C}\Bigg|_{r_1}\implies c_s^L=c_s^C=c_s \ ,
    \\
   \label{eq::smoothcond2}\frac{x'_C}{x_C}\Bigg|_{r_2}&=\frac{x'_R}{x_R}\Bigg|_{r_2}\implies c_s^R=c_s^C=c_s \ .
\end{align}
\end{subequations}
Here $c_s^L,c_s^C$ and $c_s^R$ denote the Noether charges associated with the geodesic segments in the left, central and right regions,  respectively. The matching conditions identify these charges across both branes, yielding a single conserved charge $c_s$ along the complete geodesic.

We compute the finite part of the entanglement entropy $
\log g^{(2)}_c$ for the crossing geodesic. 
A closed-form expression for \(\log g_c^{(2)}\) directly in terms of
\(\rho\) is not available. We therefore derive
\(\log g_c^{(2)}(c_s)\) and \(\rho(c_s)\), which determine this
dependence parametrically.

The expression for $\log g^{(2)}_c$ for  $r_1<0<r_2$ is computed from \eqref{eq::loggicc}.  Splitting the integral across the three wedges, this can be written as
\begin{equation}\label{eq::genericlogg2c}
    \log g^{(2)}_c= \frac{1}{4G}\left(\int^{\infty}_{r_2}dr (\mathcal{L}_R-R)+\int_{r_1}^{r_2}dr (\mathcal{L}_C- R) + \int^{r_1} _{-\infty}dr (\mathcal{L}_L-R)\right)+\log g\,,
\end{equation}
with $\mathcal{L}_{i}$ in region $i=R,C,L$ given by
\begin{equation}
  \mathcal{L}_i=\frac{R}{\sqrt{1-c_s^2 e^{-2A_i} R^{-2}}} \ .
\end{equation}
The boundary entropy $\log g$ appearing in \eqref{eq::genericlogg2c} is given in \eqref{eq::logg2TW}. Indeed, the asymptotic expansions of the exterior warp factors give
\(a_-=a\) and \(a_+=-b\), so \eqref{eq::loggapam} yields
\(\log g=\frac{c}{6}(b-a)\), in agreement with \eqref{eq::logg2TW}.

The ratio between endpoint locations is computed similarly to \eqref{eq::logratioequation} which yields 
\begin{equation}\label{eq::genericlogrho}
    \log \rho=- \int^{+\infty}_{r_2}dr \frac{c_s e^{-A_R}}{\sqrt{e^{2A_R}R^2-c_s^2}} -\int^{r_2}_{r_1}dr  \frac{c_s e^{-A_C}}{\sqrt{e^{2A_C} R^2-c_s^2}} -\int^{r_1}_{-\infty}dr \frac{c_s e^{-A_L}}{\sqrt{e^{2A_L}R^2-c_s^2}} \,.
\end{equation}
With the convention $c_s\geq0$ adopted below
\eqref{eq::diffeq}, geodesics crossing the interface satisfy
$0\leq c_s<R$. The overall minus sign in
\eqref{eq::genericlogrho} then selects the branch \(0\leq\rho\leq1\), which corresponds to  $l_L\leq l_R$, see \eqref{eq:ratiolllr}.

One might wonder whether there exist non-crossing
geodesics\footnote{Recall that non-crossing geodesics have both
endpoints on the same side of the interface.} that enter the central
AdS wedge and return to the same asymptotic boundary, as in the
single-brane examples of \cite{Anous_2022}. To reach the central wedge,
such a geodesic must first reach the brane separating it from the
exterior region, while to return to the same boundary, it must subsequently
have a radial turning point somewhere beyond that brane. Since
\(e^A\geq1\) throughout all three wedges, such a turning point requires
\(c_s>R\). However, because the exterior warp factor reaches its
minimum \(e^{A_*}=1\) between the boundary and the brane, a geodesic
with \(c_s>R\) turns before reaching the brane. Reaching the central
wedge therefore requires \(c_s<R\). These two requirements are
incompatible, so such geodesics do not exist.

Employing \eqref{eq::genericlogg2c} and using \(a=2r_1\) and \(b=2r_2\) from \eqref{eq::solIsraelJunction}, we obtain after performing the three radial integrals
\begin{equation}\label{eq::logg2cqeq1}
    \log g_c^{(2)}
=
\frac{c}{6}
\left[
-\log(\frac{R^2-c_s^2}{R^2})
-2\tanh^{-1}\!\left(
\frac{R\sinh r_1}{\mathcal{D}(r_1)}
\right)
+2\tanh^{-1}\!\left(
\frac{R\sinh r_2}{\mathcal{D}(r_2)}
\right)
\right] \ ,
\end{equation}
where we have defined
\begin{equation}
    \mathcal{D}(r)\equiv \sqrt{R^2 \cosh^2r-c_s^2} \ .
\end{equation}
Similarly, using \eqref{eq::genericlogrho} we find the ratio
\begin{equation} \label{eq::twobranesrhocs}
 \rho=
     \frac{\left(c_s-R\right) \left(c_s \sinh r_1+\mathcal{D}(r_1)\right) \left(c_s \sinh r_2-\mathcal{D}(r_2)\right)}{\left(c_s+R\right) \left(\mathcal{D}(r_1)-c_s \sinh r_1\right) \left(c_s \sinh
   r_2+\mathcal{D}(r_2)\right)} \,.
  \end{equation}
Note that both \(\rho\) and \(\log g_c^{(2)}\) are invariant under
interchanging the two tensions, \(\Sigma_1\leftrightarrow\Sigma_2\),
which sends \((r_1,r_2)\mapsto(-r_2,-r_1)\). At fixed tensions,
exchanging the two sign choices in \eqref{eq::diffeq} maps
\(\rho\to1/\rho\), while leaving \(\log g_c^{(2)}\) invariant.
Combining the two transformations reflects the whole configuration
through \(r=0\), which is an obvious symmetry.

An alternative geometric picture for the geodesics in the two-thin-wall model is presented in Appendix \ref{sec:AppendixGeometricThinBraneTwoBranes}, which provides some useful intuition.

%%%%%%%%%%%%%%%%%%%%%%%%%%%%%%%%%%%%%%%%%%%%%%%%
\subsubsection{Janus}
%%%%%%%%%%%%%%%%%%%%%%%%%%%%%%%%%%%%%%%%%%%%%%%%
For the Janus model, closed-form expressions for \(\log g^{(2)}\) as a function of \(\rho\) are not available at generic \(\beta\). We therefore turn to numerical methods. 
To this end, we first evaluate the finite contribution
$\log g^{(2)}$ and the logarithm of the endpoint ratio, $\log\rho$, as
functions of $c_s$ and the Janus parameter $\beta$. For crossing
intervals, we use \eqref{eq::loggicc} and
\eqref{eq::logratioequation}, while for non-crossing intervals, we use
\eqref{eq::loggicc2} and \eqref{eq::logratioequation2}. The parameter
$\beta$ may equivalently be traded for the boundary entropy $\log g$
using~\eqref{eq:boundentropyJ}.  
For each value of $\log g$, we treat the crossing and non-crossing configurations separately. In the crossing case, we sample the parameter range $0\leq c_s<  R\, e^{A_*}=R\sqrt{1-\frac{\beta}2}$, and construct the corresponding set of points $(\rho_c(c_s),\log g_c^{(2)}(c_s))$. Similarly, for the non-crossing configuration, we sample the regime $c_s > R \, e^{A_*}$ and construct the data set $(\rho_{nc}(c_s),\log g_{nc}^{(2)}(c_s))$. These data define the finite contribution to the entanglement entropy as a parametric function of the ratio for the crossing and non-crossing configurations.

We may partially recover analytical control by expanding to linear order in \(\beta\) the integrals \eqref{eq::logratioequation} and \eqref{eq::loggicc} for the crossing case, and \eqref{eq::logratioequation2} and \eqref{eq::loggicc2} for the non-crossing case. Although the resulting relation \(\rho=\rho(c_s,\beta)\) remains transcendental, it can be inverted perturbatively in \(\beta\) to determine \(c_s=c_s(\rho,\beta)\). We then find the following perturbative \(\log g^{(2)}\) functions for Janus:
\begin{subequations}\label{eq:expansionJanus}
\begin{align}
    \log g_c^{(2)}=&-\frac{c}6\log \frac{4 \rho }{(1+\rho )^2}+\frac{c}6\,\beta \,\left[1-\frac{\left(\rho ^2+\left(\rho ^2-4 \rho +1\right) \log \rho -1\right)}{4 \left(\rho ^2-1\right)}\right]+\mathcal{O}(\beta^2)\;,\\
    \log g_{nc}^{(2)}=&-\frac{c}{6}\log\frac{4 \rho}{(1-\rho)^2}+\frac{c}{6}\, \beta\,\left[1-\frac{ \left(\rho ^2+\left(\rho ^2+4 \rho +1\right) \log \rho -1\right)}{4 \left(\rho ^2-1\right)}\right]+\mathcal{O}(\beta^2)\,.
\end{align}
\end{subequations}
Note that for \(\beta=0\) one recovers the CFT result without interface \eqref{eq::pureAdS}. Additionally, in the limit \(\rho \to 1\), the expression for \(\log g^{(2)}_c\) given above reduces to \(\log g\), whose small-\(\beta\) expansion is \(\log g\sim\frac{c}{6}\beta\), consistently with \eqref{eq:boundentropyJ}, while \(\log g^{(2)}_{nc}\) reduces to \eqref{eq::pureAdS}, thereby recovering the CFT result, as expected since the non-crossing interval is then small compared with its distance from the interface. Furthermore, both finite contributions are invariant under the transformation \(\rho \to 1/\rho\), reflecting the symmetry of the geometry. Finally, for small values of \(\beta\), the expressions above agree with the numerical evaluation of \eqref{eq::logratioequation}, \eqref{eq::logratioequation2}, \eqref{eq::loggicc} and \eqref{eq::loggicc2} on both branches.

%%%%%%%%%%%%%%%%%%%%%%%%%%%%%%%%%%%%%%%%%%%%%%%%
\subsubsection{Super Janus}
%%%%%%%%%%%%%%%%%%%%%%%%%%%%%%%%%%%%%%%%%%%%%%%%

For the super Janus warp factor \eqref{eq:wrapSJ}, it is straightforward to  
compute analytically the finite integrals appearing in $\log g^{(2)}$, \eqref{eq::loggicc} or \eqref{eq::loggicc2}, as a function of $\rho$ 
\begin{equation}\label{eq:analyticallogg2SJ}
    \begin{aligned}
         \log g^{(2)}_{c}&=-\frac{c}{6}\left( \log \mathcal{T}+\log \frac{4 \rho^{\sqrt{{\mathcal{T}}}}}{(1+ \rho^{\sqrt{\mathcal{T}}})^2}\right) \ , \\
        \log g^{(2)}_{nc}&=-\frac{c}{6}\left( \log \mathcal{T}+ \log \frac{4 \rho^{\sqrt{{\mathcal{T}}}}}{(1- \rho^{\sqrt{\mathcal{T}}})^2}\right) \ .
    \end{aligned}
\end{equation}
The constant term is the boundary entropy $\log g$ given in \eqref{eq:boundarygSuperJanus}. Alternatively, this result, including the boundary entropy contribution, can be obtained from the empty AdS results  by the simple change of coordinates \eqref{eq:changecoordSJ}.\footnote{Note, however, that in this case in addition to applying the change of coordinates to the end-points of the interval, starting from the full entropy expression \eqref{eq:SA} together with $\log g^{(2)}$ in \eqref{eq::pureAdS}, one has to take into account 
that $\epsilon=l-l_\epsilon$ is a distance to the endpoint of the interval. Here $l$ is the location of the endpoint and $l_\epsilon$ the location where the interval is cut off. The change of coordinates maps
\begin{equation}
l\to l^{\sqrt{\mathcal{T}}}, \quad \epsilon\to l^{\sqrt{\mathcal{T}}}-l_\epsilon^{\sqrt{\mathcal{T}}} \approx \sqrt{\mathcal{T}} l^{\sqrt{\mathcal{T}}-1}\epsilon\,.
\end{equation}
} 
The result exhibits the symmetry under inversion $\rho\to 1/\rho$.
One recovers the CFT results \eqref{eq::pureAdS}  without interface for ${\cal T}=1$. One also obtains the boundary entropy \eqref{eq:boundarygSuperJanus} for a symmetric crossing interval $\log g_c^{(2)}(\rho=1)=-\frac{c}{6}\log\mathcal{T}=\log g$, as expected. On the other hand, the non-crossing finite contribution $\log g_{nc}^{(2)}$ approaches the CFT value \eqref{eq::pureAdS} when the distance of the interval to the interface is much larger than its size, $\rho\to 1$.

%%%%%%%%%%%%%%%%%%%%%%%%%%%%%%%%%%%%%%%%%%%%%%%%%%%%%%%%%%%%%%%%%%%%%%%%%%%%%%%%%%%%%%%%%%%%%%%%
\section{Entanglement transition in holographic ICFTs}\label{sec:EntanglementTransition}
%%%%%%%%%%%%%%%%%%%%%%%%%%%%%%%%%%%%%%%%%%%%%%%%%%%%%%%%%%%%%%%%%%%%%%%%%%%%%%%%%%%%%%%%%%%%%%%%

To better understand the structure of quantum correlations, one can consider
regions with more than one connected component. For example, take
$\text{A}\cup \text{B}$ in the vacuum state of a two-dimensional CFT (with no interface), with
$\text{A}=[x_1,x_2]$ and $\text{B}=[x_3,x_4]$, as illustrated in
Figure~\ref{fig:minimalsurface}. The entanglement entropy of such disconnected
regions has been studied extensively. Calabrese and Cardy originally proposed
a formula for the case of several disjoint intervals
\cite{Calabrese:2004eu}. This proposal was later shown not to be generally
valid: numerical checks and general twist-field arguments showed that, for
regions with more than one connected component, the result contains additional
theory-dependent information \cite{Caraglio:2008pk}. For two intervals, the
replica computation involves a four-point function of twist fields, whose
cross-ratio dependence is not fixed by conformal symmetry alone
\cite{Calabrese:2009ez}. Thus, unlike the single-interval vacuum entropy, the
multi-interval entropy is not determined universally by the central charge, even in the vacuum.

The entanglement entropy of disconnected regions has also been studied
holographically using the Ryu--Takayanagi prescription
\cite{Ryu_2006,Hubeny:2007re}. For two disjoint intervals, the holographic
answer is obtained by comparing competing RT configurations, as we review below. As the relative
separation of the intervals is varied, the dominant configuration can change
discontinuously, producing a sharp transition in the entanglement entropy and
in the mutual information at some critical separation \cite{Headrick:2010zt}. In this section, we analyze this transition in the different
holographic interface models introduced above.

The rest of this section is organized as follows. We first review the connected/disconnected transition for two intervals in a CFT without an interface in subsection \ref{subsec:TranNoInt}. We then present numerical results for holographic ICFTs, where the interface modifies the geodesic lengths and hence the critical separation in subsection \ref{sec:MIholoICFT}. Finally, in subsection \ref{sec:MIholoICFTSB}, we identify a universal configuration in which the interface-dependent contributions to the mutual information cancel, leaving the critical separation
at its CFT value.

%%%%%%%%%%%%%%%%%%%%%%%%%%%%%%%%%%%%%%%%%%%%%%%%
\subsection{Entanglement transition without an interface in holography }\label{subsec:TranNoInt}
%%%%%%%%%%%%%%%%%%%%%%%%%%%%%%%%%%%%%%%%%%%%%

According to the RT prescription, the entanglement entropy of the union of disconnected intervals can be obtained from geodesics in the gravity dual anchored at the endpoints of the intervals at the asymptotic boundary of space. The region enclosed by the geodesics must have the disconnected intervals as boundary, but this allows for several possible configurations, as depicted in the example of Figure \ref{fig:minimalsurface}. In this case, the entanglement entropy is determined by the configurations where the sum of the lengths of the geodesics is minimal. In higher dimensions, the geodesics become surfaces of minimal area anchored at the boundary.

For the case of two disconnected intervals, there are two geodesic configurations in the bulk that are homologous to the disconnected intervals at the boundary, as shown in Figure \ref{fig:minimalsurface}. The first, which we call the \emph{disconnected configuration}, is the union of the geodesics for the two intervals separately, see Figure \ref{fig:minimalsurface}(a), with associated geodesic length: 
\beq\label{minsur1}
\ell_{\text{disc}}\equiv\text{Length}(\gamma_{12}\cup \gamma_{34})= 2 R \ln (\frac{x_2-x_1}{\epsilon})+2 R \ln (\frac{x_4-x_3}{\epsilon}). 
\eeq
The \emph{connected configuration} instead connects \(x_1\) to \(x_4\) and
\(x_2\) to \(x_3\), as shown in Figure \ref{fig:minimalsurface}(b), and its minimal length is 
\beq\label{minsur2}
\ell_{\text{conn}}\equiv\text{Length}(\gamma_{14}\cup \gamma_{23})= 2 R \ln (\frac{x_4-x_1}{\epsilon})+2 R \ln (\frac{x_3-x_2}{\epsilon}).
\eeq
When the relative size and distance of the intervals are changed, sharp phase transitions appear between the disconnected and connected configurations of the geodesics.

\begin{figure}[htbp]
    \centering

    \begin{minipage}[t]{0.45\linewidth}
        \centering
        \includegraphics[width=\linewidth]{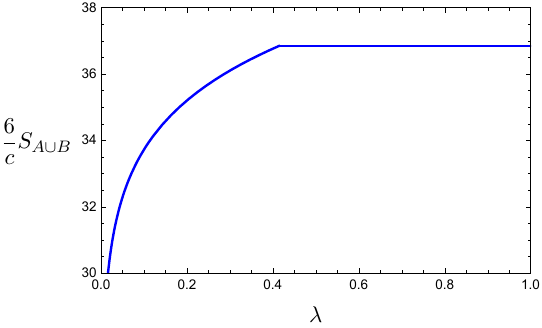}
        {\footnotesize  (a) Entropy}
    \end{minipage}
    \begin{minipage}[t]{0.52\linewidth}
        \centering
        \includegraphics[width=\linewidth]{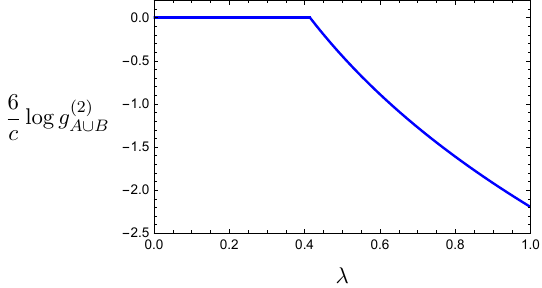}
        {\footnotesize (b) Finite part}
    \end{minipage}

    \vspace{0.5cm}

    \begin{minipage}[t]{0.52\linewidth}
        \centering
        \includegraphics[width=\linewidth]{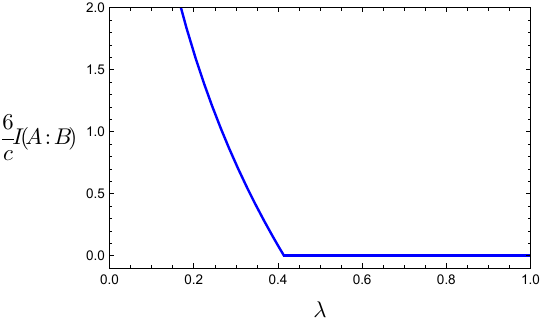}
        {\footnotesize (c) Mutual information}
    \end{minipage}
    \caption{
Entanglement transition for the union $\text{A}\cup \text{B}$ of two equal-size intervals, arranged symmetrically around the origin as in equation \eqref{eq:symadsentpts}, in pure AdS$_3$ as a function of $\lambda=h/l$. 
Panel (a) shows the entropy $S_{\text{A}\cup \text{B}}$ of the dominant RT configuration, evaluated for $l/ \epsilon=10^{4}$. Panel (b) shows the corresponding finite contribution $\log g^{(2)}_{\text{A}\cup \text{B}}$, obtained by subtracting the divergent contributions from the geodesic lengths and using the $\log g^{(2)}$ prescription \eqref{eq:SAUB}. Panel (c) shows the mutual information $I(\text{A}:\text{B})$. The transition occurs at $\lambda_C^{\text{CFT}}=\sqrt{2}-1$, where the dominant RT surface changes from the connected to the disconnected configuration. }
\label{fig:SPhCFT}
\end{figure}

Let us take two equal-size intervals of size $l$ at a distance $h$ from each other: 
\begin{equation}\label{eq:symadsentpts}
    x_1=-\frac{h}{2}-l, 
    \qquad 
    x_2=-\frac{h}{2} , 
    \qquad 
    x_3=\frac{h}{2}, 
    \qquad 
    x_4=\frac{h}{2}+l.
\end{equation}
We can plot the minimal geodesic length as a function of the distance $h$ for a size of the interval $l$ fixed, see  Figure \ref{fig:SPhCFT}(a). For short distances, the connected configuration \eqref{minsur2} has smaller length than the disconnected configuration \eqref{minsur1} and is favored. As one increases the distance between the intervals, there exists a point where both configurations have the same length, marking the transition from one type of configuration to the other. For longer distances, the disconnected configuration dominates and the entropy becomes constant. The entropy is continuous through the transition but its derivative with respect to the distance is not.

Subtracting the divergent contributions from the geodesic lengths as in
\eqref{eq:SA}, one is left with a finite regularized length. We define the finite contribution
\(\log g_{\text{A}\cup \text{B}}^{(2)}\) through the decomposition 
\begin{equation}
S_{\text{A}\cup \text{B}}
=
\frac{\min(\ell_{\text{conn}},\ell_{\text{disc}})}{4G}
\equiv
\frac{c}{6}\log\frac{2l_L^\text{A}}{\epsilon}
+
\frac{c}{6}\log\frac{2l_R^\text{A}}{\epsilon}
+
\frac{c}{6}\log\frac{2l_L^\text{B}}{\epsilon}
+
\frac{c}{6}\log\frac{2l_R^\text{B}}{\epsilon}
+
\log g_{\text{A}\cup \text{B}}^{(2)} .
\label{eq:SAUB}
\end{equation}
Here $l_L^\text{A},l_R^\text{A}$ and $l_L^\text{B},l_R^\text{B}$ are the positive distances of the
endpoints of $\text{A}$ and $\text{B}$, respectively, from the reference point $x=0$,
which we choose at the midpoint of the configuration. The full entropy is
invariant under rigid translations, although its decomposition in
\eqref{eq:SAUB} depends on this choice of reference point. The resulting
finite contribution for the dominant RT configuration is depicted in
Figure~\ref{fig:SPhCFT}(b). It vanishes for the connected configuration $\log g_{\text{A}\cup \text{B},\text{conn}}^{(2)}=0$,
since both geodesics connect pairs of endpoints placed symmetrically around
the origin. For the disconnected configuration it is given by the sum of the two non-crossing contributions in \eqref{eq::pureAdS}, namely
\begin{equation}\label{eq:discg2empty}
\log g_{\text{A}\cup \text{B},\text{disc}}^{(2)}
=-\frac{c}{3}\log\left[\lambda(\lambda+2)\right], \qquad \lambda\equiv h/l,
\end{equation}
where we have defined the dimensionless ratio $\lambda$.

A convenient way to characterize the entanglement transition is through the mutual information of the two intervals. The mutual information (MI) for two disjoint intervals A and B is cutoff-independent and given by \cite{Hayden:2011ag}
\begin{equation}\label{eq::MIbipartite}
    I(\text{A}:\text{B})=S(\text{A})+S(\text{B})-S(\text{A}\cup \text{B}) \ .
\end{equation}
An important property of the MI is its non-negativity, which follows directly from subadditivity. Moreover, when the regions A and B are disjoint, the UV divergent contributions cancel in \eqref{eq::MIbipartite}, rendering it finite. When the disconnected configuration has the smallest geodesic length, the mutual information vanishes identically. On the other hand, if the connected configuration has the smallest length, the mutual information is positive.  For the connected configuration, the mutual information is
\begin{equation}\label{eq:MutualInfconn}
    I_{\text{conn}}(\text{A}:\text{B})=\frac{c}{3}\ln\left(\frac{(x_2-x_1)(x_4-x_3)}{(x_4-x_1)(x_3-x_2)}\right).
\end{equation}
Note that the argument of the logarithm is an equal-time conformal cross-ratio.

The entanglement transition happens when  \eqref{eq:MutualInfconn} vanishes: 
\begin{equation}\label{eq:criticalCFT}
    \frac{(x_2-x_1)(x_4-x_3)}{(x_4-x_1)(x_3-x_2)}=1 \quad \Rightarrow\quad \lambda=\sqrt{2}-1\equiv\lambda_C^{\text{CFT}}.
\end{equation}
As shown in Figure \ref{fig:SPhCFT}(c), for $\lambda<\lambda_C^{\text{CFT}}$ the connected configuration is dominant and the mutual information is positive, while for $\lambda>\lambda_C^{\text{CFT}}$ the disconnected configuration is dominant and the mutual information vanishes.

Comparing the plots shown in Figure~\ref{fig:SPhCFT}, we observe that,
depending on the quantity, they become constant in different ranges of
the ratio $\lambda$. In the disconnected phase, the entropy becomes
independent of the separation, giving the plateau in
Figure~\ref{fig:SPhCFT}(a). The
$\lambda$-dependent behavior in
Figure~\ref{fig:SPhCFT}(b) for large-$\lambda$ is a consequence of the definition of $\log g_{\text{A}\cup \text{B}}^{(2)}$
in~\eqref{eq:SAUB}: its $\lambda$-dependence cancels that of the
endpoint logarithms, leaving the full entropy constant.
Even for a generic single interval in pure $\mathrm{AdS}_3$,  $\log g^{(2)}$ generally contains  ratio-dependent terms as shown in \eqref{eq::pureAdS}.
Conversely, comparison with Figure~\ref{fig:SPhCFT}(a)
makes clear that the flat part of $\log g_{\text{A}\cup \text{B}}^{(2)}$ before the
transition should not be interpreted as the full entropy being independent of the separation. Instead, in this
phase, the $\lambda$-dependence of the full entropy is carried by the
endpoint logarithms in~\eqref{eq:SAUB}.

%%%%%%%%%%%%%%%%%%%%%%%%%%%%%%%%%%%%%%%%%%%%%%%%
\subsection{Entanglement transition in holographic interfaces}\label{sec:MIholoICFT}
%%%%%%%%%%%%%%%%%%%%%%%%%%%%%%%%%%%%%%%%%%%%%%%%

In the presence of an interface, the holographic calculation of the entanglement entropy of a disconnected region proceeds in the same way as explained above. The interface modifies the geometry and changes the length of geodesics depending on their relative position with respect to the interface, which in turn modifies the location of the transition between the disconnected and connected configurations of geodesics.

\begin{figure}[htbp]
    \centering
    \includegraphics[width=0.7\linewidth]{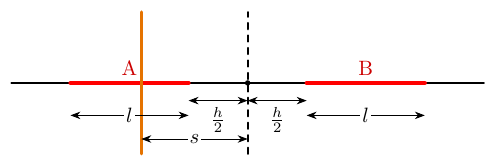}
    \caption{Setup for two disjoint boundary intervals A and B of equal length $l$, separated by a distance $h$. The orange vertical line marks the interface, taken to be at the origin, while the dashed line marks the midpoint between the two intervals. The parameter $s$ denotes the coordinate displacement of this midpoint from the interface. The figure shows the case $s>0$.}
    \label{fig:bipartitesetup}
\end{figure}

 In most cases we study a setup with equal-length intervals heuristically depicted in Figure \ref{fig:bipartitesetup}.
The length of the intervals A and B 
is $l$, the separation between them is $h$ and 
the midpoint between the two intervals is located at coordinate $s$, with the interface placed at the origin.  We use the same labels for the endpoints of the intervals as before, see Figure \ref{fig:minimalsurface}. The different (signed) positions of the different endpoints read: 
\begin{equation}\label{eq:endpointsshl}
  x_1=s -\frac{h}{2}-l, 
  \qquad x_2=s-\frac{h}{2},
  \qquad x_3= s+\frac{h}{2},
  \qquad x_4=s+ \frac{h}{2}+l .
\end{equation}
We introduce the dimensionless variables
\begin{equation}
\lambda\equiv\frac{h}{l},
\qquad
\eta\equiv\frac{x_2}{l}
=\frac{s}{l}-\frac{\lambda}{2}.
\label{eq:lambdaeta}
\end{equation}
As in the case without an interface, we study the transition by plotting
the minimal geodesic length as a function of the dimensionless separation
$\lambda$. For the non-symmetric configurations considered below,
$\lambda$ is varied at fixed $\eta$. Equation~\eqref{eq:endpointsshl}
then gives
\begin{equation}
\frac{x_1}{l}=\eta-1,
\qquad
\frac{x_2}{l}=\eta,
\qquad
\frac{x_3}{l}=\eta+\lambda,
\qquad
\frac{x_4}{l}=\eta+\lambda+1.
\end{equation}
For $0<\eta<1$, the interface lies inside A, and $\eta$
measures the fraction of A lying to the right of the interface.\footnote{For example, $\eta=0.5$ places the interface at the midpoint of A, whereas
$\eta=0.1$ places it close to the right endpoint. We use those values later on in the plots.} 
Thus, with distances measured in units of $l$, increasing $\lambda$
at fixed $\eta$ keeps the position of the interface within A unchanged
while translating the interval B to the right and increasing the
separation between the two intervals.

\begin{figure}[htbp]
    \centering

    % Top row: Thin Wall and Janus
    \begin{minipage}[t]{0.47\linewidth}
        \centering
        \includegraphics[width=\linewidth]{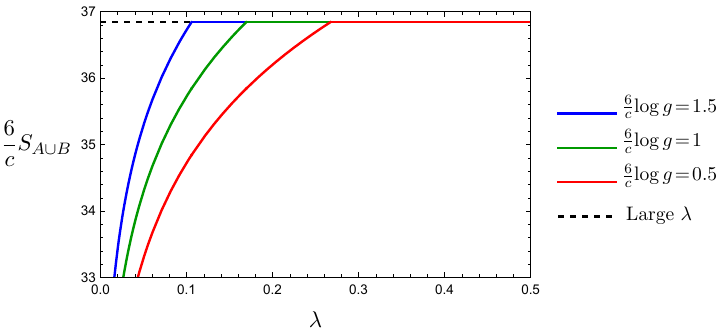}

        \small (a) Single-brane thin wall
    \end{minipage}
    \hfill
    \begin{minipage}[t]{0.47\linewidth}
        \centering
        \includegraphics[width=\linewidth]{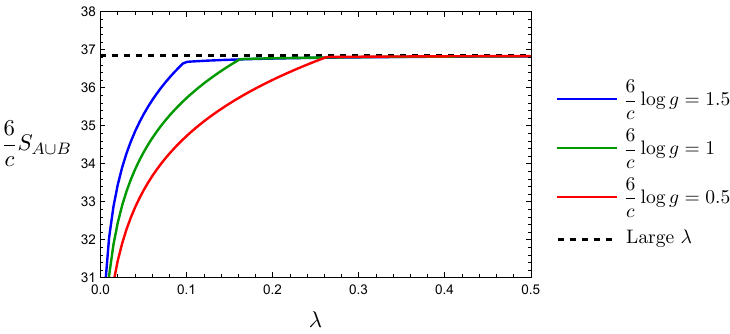}

        \small (b) Janus
    \end{minipage}

    \vspace{0.5cm}

    % Bottom row: Super Janus
    \begin{minipage}[t]{0.47\linewidth}
        \centering
        \includegraphics[width=\linewidth]{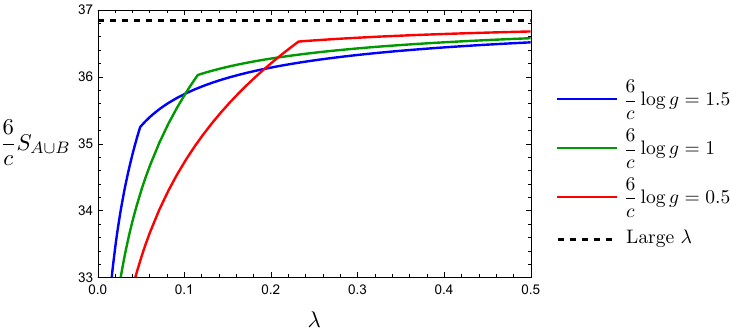}

        \small (c) Super Janus
    \end{minipage}

    \caption{Entropy transitions of intervals arranged in a symmetric configuration around the interface, for various values of $\log g$.
    The different panels correspond to the different models: (a) Single-brane thin wall, (b) Janus, and (c) Super Janus. The plots are made fixing  $l/\epsilon=10^4$.}
    \label{fig:entropy-phase-transitions}
\end{figure}

\begin{figure}[htbp]
    \centering

    % Top row: Thin Wall and Janus
    \begin{minipage}[t]{0.47\linewidth}
        \centering
        \includegraphics[width=\linewidth]{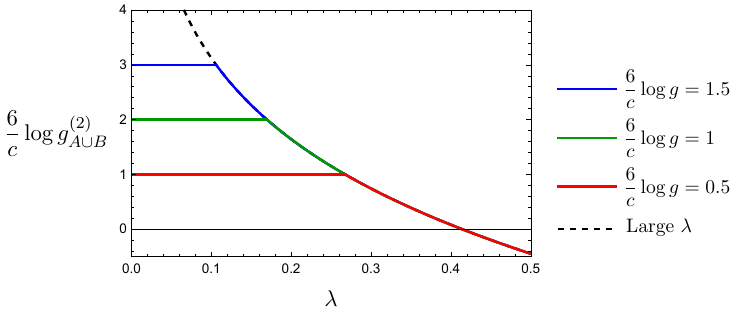}

        \small (a) Single-brane thin wall
    \end{minipage}
    \hfill
    \begin{minipage}[t]{0.47\linewidth}
        \centering
        \includegraphics[width=\linewidth]{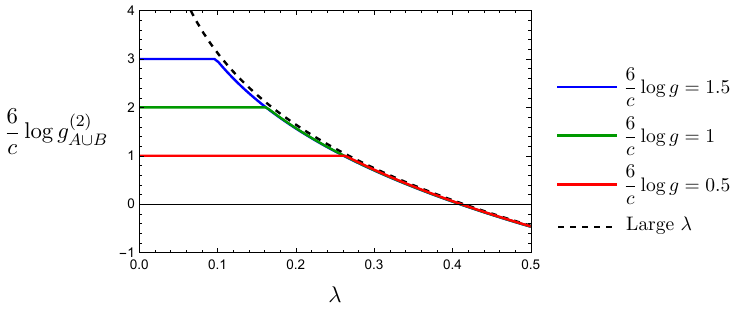}

        \small (b) Janus
    \end{minipage}

    \vspace{0.5cm}

    % Bottom row: Super Janus
    \begin{minipage}[t]{0.47\linewidth}
        \centering
        \includegraphics[width=\linewidth]{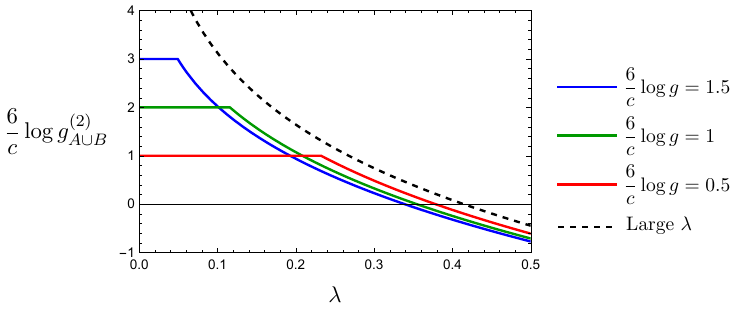}

        \small (c) Super Janus
    \end{minipage}

    \caption{Finite part of entropy transitions of intervals arranged symmetrically around the interface, for various values of $\log g$.
    (a) Single-brane thin wall, (b) Janus, and (c) Super Janus. 
    }
    \label{fig:finite-entropy-phase-transitions}
\end{figure}

For the symmetric configuration \(s=0\), Figures~\ref{fig:entropy-phase-transitions}
and \ref{fig:finite-entropy-phase-transitions} show, respectively, the full
entropy and the finite part $\log g^{(2)}_{\text{A}\cup\text{B}}$, see equation \eqref{eq:SAUB}, for the single-brane thin-wall,
Janus, and super-Janus models.
At small separations, the connected configuration has smaller geodesic length than the disconnected one and therefore is dominant. As the separation is increased, the two configurations become degenerate at a critical value of $\lambda$. This critical point appears as a kink in the minimal length curve and signals the transition from the connected to the disconnected phase. For separations beyond this point, the disconnected configuration has the smaller geodesic length and hence is favored.

For large values of $\lambda$, when the two intervals are taken far apart (and still symmetric about the interface), the finite contribution to the entanglement entropy approaches the CFT expression \eqref{eq:discg2empty}, namely 
\begin{equation}
    \log g^{(2)}_{\text{A}\cup \text{B},\text{disc}}=-\frac{c}{3}\log\!\left[\lambda(\lambda+2)\right]+\ldots \ ,\qquad \lambda\gg 1,
\end{equation}
regardless of the model and independently of $\log g$. Equivalently, the entanglement entropy should approach the disconnected plateau
\begin{equation}
S_{\text{A}\cup \text{B},\text{disc}}=
\frac{2c}{3}\log\frac{l}{\epsilon}+\ldots, \qquad \lambda\gg 1 \ .
\end{equation}
These large $\lambda$ asymptotics are depicted as black dashed curves in Figures \ref{fig:entropy-phase-transitions} and \ref{fig:finite-entropy-phase-transitions}.

Having illustrated the transition for the symmetric configuration,
we now systematically study how the critical separation depends on the
interface parameters and on the relative placement of the interface and
the intervals. For each configuration, we evaluate
$I_{\mathrm{conn}}(\text{A}:\text{B})$, namely the mutual information defined
in~\eqref{eq::MIbipartite} with $S(\text{A}\cup \text{B})$ evaluated on the connected
RT candidate, and locate the transition by solving
$I_{\mathrm{conn}}(\text{A}:\text{B})=0$. Upon substituting the single-interval
decomposition~\eqref{eq:SA}, the endpoint terms cancel pairwise in
$I_{\mathrm{conn}}(\text{A}:\text{B})$, leaving a combination of the finite
contributions $\log g^{(2)}$. For each term, we use
$\log g_c^{(2)}$ or $\log g_{nc}^{(2)}$, as obtained in
Section~\ref{sec:HolographicEE}, according to whether the corresponding
boundary interval has endpoints on opposite sides of the interface or
lies entirely on one side.

In all cases, for each interval the ratio of the distances of the
endpoints from the interface, $\rho=l_L/l_R$, and the finite part of
the entanglement entropy, $\log g^{(2)}$, are both determined by
$c_s$, once we specify whether the interval is crossing or
non-crossing and, in the latter case, on which side of the interface
it lies. Thus, solving for $c_s$ gives the corresponding
$\log g^{(2)}$ as a function of $\rho$. 
For a crossing interval, exchanging the two sign choices
in~\eqref{eq::diffeq} maps $\rho\to1/\rho$ while leaving
$\log g_c^{(2)}$ invariant.\footnote{This can also be understood directly
from defect conformal symmetry. After extracting the endpoint
dependence fixed by conformal covariance, the crossing contribution
depends only on the cross-ratio
$
\frac{4l_Ll_R}{(l_L+l_R)^2}
=\frac{4\rho}{(1+\rho)^2}$,
which is invariant under $\rho\to1/\rho$.} 
This transformation
preserves the two sides of the interface rather than exchanging
them, and therefore does not require the background itself to be
reflection symmetric. 
For a non-crossing
interval, the corresponding functions on the two sides need not
agree in a general asymmetric geometry, but they coincide in all the
models considered here.\footnote{This equality is not generic. In
the single-brane, Janus, and super-Janus geometries it follows from
reflection symmetry. In the double-brane thin-wall model with
unequal tensions, the geometry is not reflection symmetric. The
equality instead follows because the relevant non-crossing geodesics
remain entirely within the equal-radius pure-AdS exterior regions.
For a general asymmetric geometry, and in particular when the
exterior AdS radii are unequal, one should distinguish
$\log g_{nc,L}^{(2)}$ and $\log g_{nc,R}^{(2)}$. Moreover, the
dominant extremal-geodesic branch may itself depend on the side.}
Using these properties, it is sufficient here to evaluate the
$\log g^{(2)}$ functions with $0<\rho\leq1$. For a non-crossing
interval, this amounts to choosing the representative configuration
to the right of the interface, for which $l_L<l_R$. An interval to
the left is included by inverting its original ratio. For a crossing interval, this means choosing the configuration in
which the portion of the interval to the right of the interface is at
least as long as the portion to the left.  
With these identifications,
the five generic placements of the interface reduce to the following
three inequivalent cases for the models considered where the $\log g^{(2)}$ functions are restricted to $\rho\leq 1$ values:

\begin{itemize}
\item The interface is between the two intervals: $x_1<x_2<0<x_3<x_4$. Assuming for instance the hierarchy   $|x_1|\leq x_4$, $|x_2|\leq x_3$:
\begin{align}\label{eq::Icbetween}
    I_{\text{conn}}(\text{A}:\text{B})&=  \,\log^{(2)}_{nc} \left(\frac{|x_2|}{|x_1|}\right)+\log^{(2)}_{nc} \left(\frac{x_3}{x_4}\right)-\log^{(2)}_{c} \left(\frac{|x_1|}{x_4}\right)-\log^{(2)}_{c} \left(\frac{|x_2|}{x_3}\right) . 
\end{align}
If either of the endpoint hierarchies specified above is reversed,
the corresponding argument of $\log g_c^{(2)}$ should be inverted $\rho\rightarrow 1/\rho$.
\item Both intervals are on the same side of the interface. If we take them to be to the right, $0<x_1<x_2<x_3<x_4$: 
\begin{align}\label{eq::ICoutside}
    I_{\text{conn}}(\text{A}:\text{B})&=  \,\log^{(2)}_{nc} \left(\frac{x_1}{x_2}\right)+\log^{(2)}_{nc} \left(\frac{x_3}{x_4}\right)-\log^{(2)}_{nc} \left(\frac{x_1}{x_4}\right)-\log^{(2)}_{nc} \left(\frac{x_2}{x_3}\right)  .
\end{align}
If both intervals lie to the left of the interface, the same
expression applies with each argument of $\log g_{nc}^{(2)}$
inverted, $\rho\to1/\rho$, so that all arguments again lie in
$0<\rho<1$. 
\item The interface is inside one of the intervals. Take for example the interface inside A, $x_1<0<x_2<x_3<x_4$, for the case where $x_2\leq|x_1|<x_4$: 
\begin{align}\label{eq:Icinside}
    I_{\text{conn}}(\text{A}:\text{B})&=  \,\log g^{(2)}_{c} \left(\frac{x_2}{|x_1|}\right)+\log g^{(2)}_{nc} \left(\frac{x_3}{x_4}\right)-\log g^{(2)}_{c} \left(\frac{|x_1|}{x_4}\right)-\log g^{(2)}_{nc} \left(\frac{x_2}{x_3}\right) . 
\end{align}
If instead $|x_1|<x_2$, the  argument of the first
$\log g_c^{(2)}$ should be inverted, while if $|x_1|>x_4$, the
 argument of the second $\log g_c^{(2)}$ should be inverted. 
The case in which the interface lies inside B is obtained from
Eq.~\eqref{eq:Icinside} by reflecting the configuration and relabelling
the ordered endpoints according to
$
 (x_1,x_2,x_3,x_4)\longrightarrow
 (-x_4,-x_3,-x_2,-x_1)$.
This maps \(x_1<0<x_2<x_3<x_4\) to
\(x_1<x_2<x_3<0<x_4\), exchanges the roles of A and B, and
maps the hierarchy \(x_2\leq|x_1|<x_4\) to
\(|x_3|\leq x_4<|x_1|\). Substituting this map into
Eq.~\eqref{eq:Icinside} and writing the left-hand endpoint distances
as \(|x_i|\) gives the corresponding expression with all arguments in
\(0<\rho\leq1\).
For other endpoint hierarchies, the corresponding
\(\log g_c^{(2)}\) arguments are inverted as described above, whenever
necessary to keep them in the range \(0<\rho\leq1\). 
\end{itemize}

With these expressions at hand, we now determine the critical values of $\lambda$ where the mutual information vanishes,  for different values of the boundary entropy, namely  $\lambda_C(\log g)$. In the absence of an interface, the critical separation is given
by~\eqref{eq:criticalCFT}. As before, for values of $\lambda<\lambda_C$, the connected configuration is dominant and the mutual information is positive, whereas for $\lambda>\lambda_C$ the disconnected configuration is favored and the mutual information vanishes.

\begin{figure}[htbp]
    \centering

    % Top row
    \begin{minipage}[t]{0.47\linewidth}
        \centering
        \includegraphics[width=\linewidth]{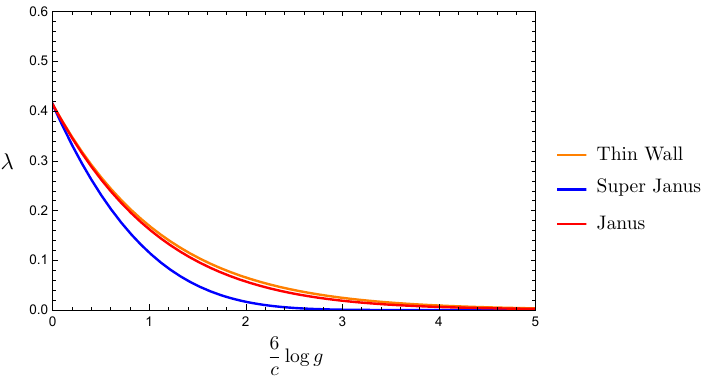}

        \small (a) Symmetric setup
    \end{minipage}
    \hfill
    \begin{minipage}[t]{0.47\linewidth}
        \centering
        \includegraphics[width=\linewidth]{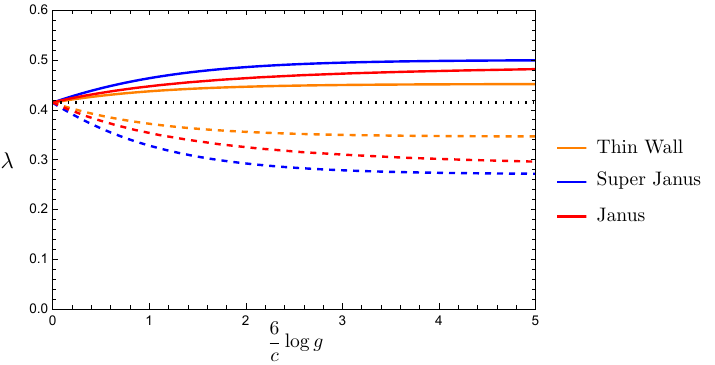}

        \small (b) Interface inside $\text{A}$
    \end{minipage}

    \vspace{0.5cm}

    % Bottom row
    \begin{minipage}[t]{0.47\linewidth}
        \centering
        \includegraphics[width=\linewidth]{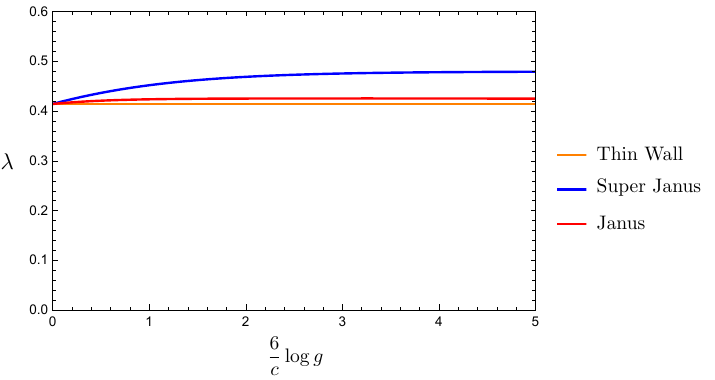}

        \small (c) Interface outside $\text{A}$
    \end{minipage}
\caption{Critical curves for the three models (thin-wall, Janus and super Janus) for different interface positions:
(a) fully symmetric setup,
(b) interface inside A, with $\eta=0.5$ (solid curves)
and $\eta=0.1$ (dashed curves), and
(c) both intervals to the right of the interface, with $\eta=1.1$. See equation \eqref{eq:lambdaeta} for the definition of $\eta$. The horizontal black dotted line in (b) denotes the no-interface value
in~\eqref{eq:criticalCFT}
}
    \label{fig:critical-curves}
\end{figure}

We first apply this procedure to the single-brane thin wall, Janus and Super Janus models. The resulting critical curves in the $(\lambda,\log g)$ plane for the different interface configurations in each of the three models are shown in Figure \ref{fig:critical-curves}. 
To determine the critical curves, we employ the following numerical procedure. For the single-brane thin wall and super Janus models we use the analytic formulas for $\log g^{(2)}$ as a function of the ratio $\rho$, while for the Janus model we use the numerical methods discussed in Section \ref{sec:formulasModels}. Using these functions, we evaluate the mutual information for the three possible positions of the interface: interface lying between the two intervals \eqref{eq::Icbetween}, outside both intervals \eqref{eq::ICoutside}, or inside one of them \eqref{eq:Icinside}.

For each fixed value of $\log g$, we vary $\lambda$ along the
one-parameter family associated with each panel: we fix $s=0$ for
Figure~\ref{fig:critical-curves}(a), while fixing $\eta$, as defined
in~\eqref{eq:lambdaeta}, to $0.5$ or $0.1$ for
Figure~\ref{fig:critical-curves}(b), and to $1.1$ for
Figure~\ref{fig:critical-curves}(c). We then numerically determine the
critical separation $\lambda_C(\log g)$ from
$I_{\mathrm{conn}}\bigl(\lambda_C,\log g\bigr)=0$.
Repeating this procedure for a range of values of $\log g$ yields the
critical curves shown in Figure~\ref{fig:critical-curves}.

For the single-brane thin-wall model,
Appendix~\ref{sec:AppendixGeometricThinBraneTwoIntervals} provides an
alternative geometric derivation of the transition conditions with the
interface between the two intervals or inside one of them, using the
elementary geometry of circular RT geodesic arcs in Poincaré coordinates.

\begin{figure}[htbp]
    \centering

    \begin{minipage}[t]{0.6\linewidth}
        \centering
        \includegraphics[width=\linewidth]{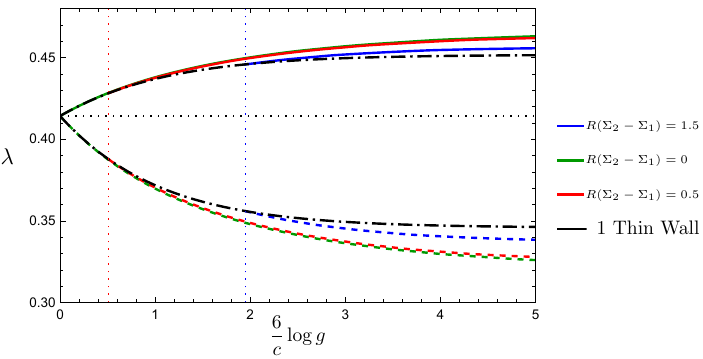}

    \end{minipage}
    \caption{Critical curves for the double thin-wall model with the interface located inside $\text{A}$, for equal AdS radii in the three wedges and several values of the difference in brane tensions $R(\Sigma_2-\Sigma_1)$. 
At each fixed value of $R(\Sigma_2-\Sigma_1)$, the solid and dashed
curves correspond to $\eta=0.5$ and $\eta=0.1$, respectively.  
Positive subcritical brane tensions require $-2< R(\Sigma_2-\Sigma_1) < 2$. 
At fixed \(R(\Sigma_2-\Sigma_1)\), positivity of both brane tensions
imposes the lower bound on \(\log g\) given in
\eqref{eq::boundaryentrbound}, indicated by the vertical dotted lines.  
The horizontal black dotted line denotes the no-interface value
in~\eqref{eq:criticalCFT}, while the two non-horizontal dot-dashed black curves
show the corresponding single-brane thin-wall results.
}
    \label{fig:ccdb}
\end{figure}

Next, we extend the analysis of the critical separation to the
double-brane thin-wall model, using the results of
Subsection~\ref{sec:twoTW}, and show in
Figure~\ref{fig:ccdb} how the critical curves for the
interface-inside-A configuration depend on the additional independent
parameter $R(\Sigma_2-\Sigma_1)$.

More generally, the different models exhibit the following common
qualitative behaviors. When the interface is between the two intervals, the critical ratio $\lambda_C$ has a value smaller than the critical value without interface, $\lambda_C^{\text{CFT}}$, given in \eqref{eq:criticalCFT}, and it is monotonically decreasing with $\log g$. On the other hand, if the two intervals are on the same side of the interface, the behavior is the opposite, $\lambda_C$ is larger or equal in the presence of the interface, and it is non-decreasing with $\log g$. This suggests that bipartite entanglement is redistributed by the presence of the interface, decreasing for intervals separated by the interface and increasing or not changing, when the intervals happen to be on the same side.

In the large-$\log g$ limit, where the transmission coefficient
approaches its minimum value, $\lambda_C\to0$ when the interface lies
between the two intervals. Otherwise, $\lambda_C$ approaches a finite
value that generally depends on $\eta$ and on the model. When the
interface lies outside both intervals, this limiting value satisfies
$\lambda_C\geq\lambda_C^{\mathrm{CFT}}$, with equality for the
thin-wall models. 
These asymptotes can also be obtained by substituting the
large-$\log g$ asymptotic forms of $\log g_c^{(2)}(\rho)$ and
$\log g_{nc}^{(2)}(\rho)$ into the appropriate mutual-information
expression, \eqref{eq::Icbetween}, \eqref{eq::ICoutside}, or
\eqref{eq:Icinside}, and solving $I_{\mathrm{conn}}(\text{A}:\text{B})=0$ at leading
nonvanishing order. We have checked that the resulting values are consistent with the
numerical curves in Figures~\ref{fig:critical-curves}
and~\ref{fig:ccdb}.

Interestingly, for the interface-inside-A configuration \eqref{eq:Icinside}, analyzed
in Figures~\ref{fig:critical-curves}(b) and~\ref{fig:ccdb}, there is a  crossover as the location of the interface is changed along the interval, interpolating between the qualitative behaviours observed for intervals on the same side and on opposite sides of the interface. This is reflected in the
opposite monotonicities of the solid and dashed curves.

This suggests that there is an intermediate position of
the interface at which the critical curve is stationary at $\log g=0$.
This position can be determined by expanding the mutual information of the connected configuration for
small $\log g$, finding the critical value where it vanishes, $\lambda_C(\log g,\eta)$, and solving
\begin{equation}
\left.
\frac{\partial\lambda_C(\log g;\eta)}
{\partial\log g}
\right|_{\log g=0}=0
\end{equation}
for $\eta$. For the single-brane thin-wall and super-Janus models, this
condition can be evaluated using the analytic formulas for the finite
contribution to the entanglement entropy derived in
Section~\ref{sec:formulasModels}. For the Janus solution, one can instead
use the expansion in~\eqref{eq:expansionJanus}. In all three cases, the
calculation gives \footnote{The label refers to ``Silver Blaze'', as explained in the Introduction.}
\begin{equation}\label{eq:etaSB}
\lambda_{SB}=\lambda_C^{\mathrm{CFT}}=\sqrt{2}-1,\qquad \eta_{SB}=1-\frac{1}{\sqrt{2}}.
\end{equation}
The following subsection establishes the stronger result that, at 
$\eta=\eta_{SB}$, an exact, model-independent cancellation in the mutual
information gives $\lambda_C=\sqrt{2}-1$ for every allowed value of $\log g$,
making the critical curve exactly horizontal. Consequently, the same
result holds for every interface model studied here, including the
double-brane thin-wall model.

%%%%%%%%%%%%%%%%%%%%%%%%%%%%%%%%%%%%%%%%%%%%%%%%
\subsection{Critical values and the universal Silver Blaze configuration}\label{sec:MIholoICFTSB}
%%%%%%%%%%%%%%%%%%%%%%%%%%%%%%%%%%%%%%%%%%%%%%%%

Let us first recall the expression for the mutual information of the connected geodesic configuration when the interface is inside one of the intervals, given by equation \eqref{eq:Icinside}. We observe that there are contributions from both crossing and non-crossing intervals that contribute with opposite signs. If we impose that the arguments of each type of contribution are the same, there will be an exact cancellation and the mutual information will vanish. When the interface is inside the interval A, this leads to the conditions 
\begin{equation}
    \frac{x_2}{|x_1|}=\frac{|x_1|}{x_4},\quad \frac{x_3}{x_4}=\frac{x_2}{x_3}.
\end{equation}
These are equivalent to the following conditions on the endpoints of the intervals
\begin{equation}\label{eq:silverblazepoint}
    -x_1=x_3,\qquad  x_3=\sqrt{x_2 x_4}.
\end{equation}

Geometrically, in the configuration \eqref{eq:silverblazepoint}, the left endpoints of the intervals are related by a reflection through the interface.  They are also localized at a distance from the interface that is the geometric mean of the right endpoints. With these conditions, \eqref{eq:Icinside} becomes 
\begin{align}\label{eq:silverblazecancelation}
    I_{\text{conn}}(\text{A}:\text{B})&=  \,\log g^{(2)}_{c} \left(\sqrt{\frac{x_2}{x_4}}\right)+\log g^{(2)}_{nc} \left(\sqrt{\frac{x_2}{x_4}}\right)-\log g^{(2)}_{c} \left(\sqrt{\frac{x_2}{x_4}}\right)-\log g^{(2)}_{nc} \left(\sqrt{\frac{x_2}{x_4}}\right)=0 .
\end{align}
Thus, this arrangement corresponds to a critical configuration for any semiclassical two-dimensional holographic ICFT with equal central
charges on the two sides. There is an analogous configuration when the interface is inside the interval B.

The cancellation in equation \eqref{eq:silverblazecancelation} follows directly from the scale invariance of the interface problem. After separating the divergent endpoint logarithms as in equation \eqref{eq:SA}, the remaining finite single-interval contribution depends only on the ratio of endpoint distances from the interface and on whether the interval is crossing or non-crossing. 
The divergent CFT endpoint logarithms cancel in the mutual information, leaving 
exactly the sum of  terms in equation \eqref{eq:Icinside}. 
Then imposing the condition \eqref{eq:silverblazepoint} makes the cancellation in equation \eqref{eq:silverblazecancelation}
automatic: the two crossing terms have the same argument and opposite signs,
as do the two non-crossing terms. This explains why equation \eqref{eq:silverblazecancelation}
is universal and not model-specific.

Let us now generalize our setup by allowing the intervals A and B to have different lengths given by $l_{\text{A}}$ and $l_{\text{B}}$ respectively. The positions of the endpoints can be written as
\begin{equation} \label{eq::kappasetup}
    x_1=-l_{\text{A}}+s-\frac{h}{2} \ ,\quad x_2=s-\frac{h}{2} \ ,\quad x_3=s+\frac{h}{2} \ , \quad x_4=l_{\text{B}}+s+\frac{h}{2} \ .
\end{equation}
 We redefine the dimensionless ratios as
 \begin{equation}\label{eq:SBgeneral1}
     \lambda=\frac{h}{l_{\text{A}}},\quad \sigma=\frac{s}{l_{\text{A}}},\quad \kappa=\frac{l_{\text{B}}}{l_{\text{A}}}\,.
 \end{equation}
The conditions \eqref{eq:silverblazepoint} translate into
\begin{equation}\label{eq:SBgeneral2}
    \sigma_{SB}=\frac{1}{2},\qquad \lambda_{SB}=\frac{1}{2} \left(\sqrt{\kappa ^2+6 \kappa +1}-\kappa -1\right)\,.
\end{equation}
For intervals of the same length, $\kappa=1$, we recover $\lambda_{SB}=\sqrt{2}-1=\lambda_C^{\text{CFT}}$. In this case, equation~\eqref{eq:lambdaeta} gives
$\eta_{\mathrm{SB}}
=\sigma_{\mathrm{SB}}-\lambda_{\mathrm{SB}}/2
=1-1/\sqrt{2}$, in agreement with the result above in equation~\eqref{eq:etaSB}. More generally, since the Silver Blaze result does not depend on the interface, $\lambda_{\mathrm{SB}}$  also gives the critical value of $\lambda$ for a CFT without an interface for arbitrary $\kappa$. 

An alternative geometric derivation of the Silver Blaze configuration along with some additional details on the entropy behavior at this point can be found in appendix \ref{app:subappSB}.

%%%%%%%%%%%%%%%%%%%%%%%%%%%%%%%%%%%%%%%%%%%%%%%%%%%%%%%%%%%%%%%%%%%%%%%%%%%%%%%%%%%%%%%%%%%%%%%%
\section{Discussion}\label{sec:Discussion}
%%%%%%%%%%%%%%%%%%%%%%%%%%%%%%%%%%%%%%%%%%%%%%%%%%%%%%%%%%%%%%%%%%%%%%%%%%%%%%%%%%%%%%%%%%%%%%%%

In this paper, we computed the interface contribution to the finite part of the entanglement entropy, $\log g^{(2)}$, for several families of holographic models dual to two-dimensional interface CFTs, including thin-wall geometries with one or two constant-tension branes, as well as Janus and super-Janus solutions. This has enabled us to investigate in detail how the entanglement transition for configurations of two intervals is affected by the presence of the interface.

We derived explicit analytic expressions for $\log g^{(2)}$ in a subset of the examples and obtained numerical results for the remaining cases. Using these results, we studied entanglement transitions for two intervals, which in the holographic description correspond to transitions between connected and disconnected configurations of geodesics anchored at the  endpoints of the intervals. We employed the mutual information between the intervals as a diagnostic of the transition. The transition occurs at critical configurations where the mutual information changes from zero to non-zero. For intervals of fixed and equal length, we determined the critical separation as a function of the boundary entropy $\log g$ for each class of models. We systematically explored all possible relative positions of the intervals with respect to the interface.

For intervals placed symmetrically around the interface, we found that the critical separation depends sensitively on $\log g$. As $\log g$ increases, the transition occurs at smaller separations, indicating weaker correlations between the two intervals. In the limiting case $\log g \to \infty$, the transmission through the interface becomes minimal and the mutual information of the two intervals vanishes regardless of their separation. Conversely, as $\log g \to 0$, the critical separation approaches the value obtained in a CFT without an interface.

As the two-interval configuration is translated relative to the
interface, so that the interface moves from the gap between the
intervals, through one of them, and eventually outside both, the
critical separation initially continues to decrease with $\log g$, but
this trend changes as the interface passes through the interval.
In particular, for the Janus and super-Janus models,
when both intervals lie on the same side of the interface, the
dependence on $\log g$ is opposite: increasing $\log g$ leads to a
larger critical separation. If the interface is sufficiently far away,
the intervals cease to be sensitive to its presence, and the transition
occurs at the same critical separation as in the absence of an
interface. In the equal AdS-radius thin-wall models studied here, this regime is reached as soon as both
intervals are located on the same side of the interface.

 The crossover from the reduced critical separation found when the
interface lies between the intervals to the enhanced-or-unchanged
critical separation found when both intervals lie on the same side
occurs as the interface passes through one of the intervals. There
exists a special position within that interval, common to all the
models studied here, at which the critical separation coincides with
the no-interface value throughout the allowed range of $\log g$. We
refer to this setup as the \emph{Silver Blaze configuration}. In the semiclassical approximation of the gravity dual, this cancellation is
universal for conformal holographic interfaces in two-dimensional
CFTs with equal central charges on the two sides (although, as discussed in the following paragraph, we expect it to
persist also for unequal central charges).

This work can be extended in several directions.  
All the models studied in this paper have equal central charges on the two sides of the interface, corresponding
to equal asymptotic values of the AdS radii near the two boundaries in
the dual description.\footnote{This does not imply reflection symmetry: the unequal-tension
double-brane thin-wall background is generally asymmetric, although,
as a special property of the equal-radius construction, the
single-interval entropy functions entering our analysis, and hence the
resulting entanglement transitions, are invariant under exchanging the
two brane tensions, see Subsection~\ref{sec:twoTW} and
Appendix~\ref{sec:AppendixGeometricThinBraneTwoBranes}.} 
It is interesting to ask what happens when we allow two different central charges on the two sides of the interface. In other contexts, such as the study of energy transmission this induces  qualitative differences compared to the equal central charge scenario, for instance full transmission from the side with the larger central charge becomes impossible \cite{Meineri_2020,Bachas_2020}. 
Allowing for unequal central charges is expected to modify the entanglement transition too, especially  for intervals located on opposite sides of the interface. 
We nevertheless expect the Silver Blaze cancellation
to persist: at the Silver Blaze point, the two crossing contributions
are evaluated at the same ratio and cancel, as do the two non-crossing
contributions on the same side, even though the finite entropy
functions may differ between the two sides. This expectation should be
tested in explicit unequal-radius backgrounds. In thin-wall models, introducing different central charges is straightforward, as it simply requires geometries with distinct AdS radii on either side of the interface. The extension to Janus-type solutions like the ones constructed in \cite{Arav:2020asu,Chen:2021mtn,Gutperle:2022fma,Ghodsi:2022umc} is more challenging, since simple analytic solutions are not known. 
Nevertheless, such constructions might in principle be accessible using the fake supergravity formalism of \cite{Freedman:2003ax}. We will address these cases in future work. Other possible extensions include multiway junctions connecting more than two spatial regions at the interface \cite{Shen:2024itl,Chakraborty:2025jtj,Chakraborty:2026wip}.

The Silver Blaze configuration relies both on the geometric constraints imposed by the low dimensionality of the theory and on the scaling symmetries of the interfaces. In higher dimensions, the two-interval configuration studied here is naturally generalized to two strips whose boundaries are parallel to the interface. For holographic CFTs, the critical ratio between the strip separation and width at which the entanglement transition occurs depends on the spacetime dimensionality, e.g., \cite{Alishahiha:2014jxa,Ben-Ami:2014gsa,Balasubramanian:2018qqx}. 
It would therefore be interesting to investigate whether Silver Blaze configurations persist in higher-dimensional holographic ICFTs, whether their critical ratios coincide with the corresponding
no-interface values, and whether analogous distinguished configurations arise  
for other shapes of entangling regions.

Breaking conformal invariance may give rise to new phenomena in the
entanglement transition, as has been observed in theories with a mass
gap~\cite{Jokela:2020wgs}, holographic RG
flows~\cite{Balasubramanian:2018qqx}, and non-zero temperature and
density states~\cite{Tonni:2010pv,deOliveira:2025qwe,Kundu:2016dyk}. Related holographic interface and boundary settings include finite-temperature solutions~\cite{Bak:2011ga,Bak:2013uaa,
Estes:2015jha,Bachas:2021fqo,Bachas:2021tnp},
defect-localized RG flows~\cite{Erdmenger:2013dpa,
Erdmenger:2015spo,Erdmenger:2020hug,Kanda:2023zse},
and 
interfaces connecting theories related by an RG
flow, commonly referred to as RG interfaces~\cite{
Melby-Thompson:2017aip,Gutperle:2012hy,Arav:2020asu,
Chen:2021mtn,Gutperle:2022fma,Giombi:2024qbm,
Gutperle:2024yiz,Karndumri:2025dqe,Afxonidis:2026txc}.

Beyond entanglement entropy and mutual information, several related correlation
measures could be studied in holographic ICFTs. One natural extension is to consider
multipartite information and other multi-region quantities built from entropies of more
than two intervals. In holography these are obtained by the same RT prescription, by
comparing the relevant competing extremal-surface configurations for multiple boundary
regions. Such multipartite quantities are known to exhibit phase transitions in holographic
theories \cite{Hayden:2011ag,Alishahiha:2014jxa,Ben-Ami:2014gsa,Mirabi:2016elb}.
It would be interesting to understand how the interface modifies these multi-region transitions.

Rényi entropies provide another direction, but they are more subtle.
For a single ball-shaped region in a CFT (a
single interval in two dimensions), they can be computed by a
conformal mapping to a thermal problem on hyperbolic
space~\cite{Hung:2011nu}. In the presence of an interface, this mapping remains directly applicable only for special
placements of the interval relative to the interface, such as when the
interval is centered on the interface \cite{Jensen:2013lxa}. A more general holographic
approach is Dong's cosmic-brane prescription, in which the Rényi
entropy is computed from the area of a codimension-two brane whose
backreaction modifies the bulk geometry~\cite{Dong:2016fnf}. For the two-interval configurations considered here, the problem
at Rényi index $n\neq1$ is already nontrivial even in the vacuum
without an interface. Unlike at $n=1$, it cannot in general be reduced
to independent single-interval results. Existing treatments include a
perturbative analysis around $n=1$ using the cosmic-brane
prescription~\cite{Dong:2016fnf}, and a numerical analysis at general $n$ using
the gravitational replica construction~\cite{Faulkner:2013yia}.
Applying the cosmic-brane prescription to holographic ICFTs would additionally require solving the backreacted bulk geometry in the
presence of the interface. In the thin-wall models, the interface-brane
action and the associated junction conditions must also be included.

Another holographic probe of correlations between subregions is the
entanglement wedge cross section (EWCS), which has been evaluated
in semiclassical holographic ICFTs~\cite{Kusuki:2022bic,Tang:2023chv,
Basak:2023bnc}. The EWCS was first proposed as
the holographic dual of entanglement of
purification~\cite{Takayanagi:2017knl,Nguyen:2017yqw}, and was later
related to reflected entropy~\cite{Dutta:2019gen}. The EWCS satisfies
$E_W(\text{A}:\text{B})\geq\frac{1}{2}I(\text{A}:\text{B})$, and, in the semiclassical limit of the gravity dual, the reflected entropy is related to the
EWCS by 
$S_R(\text{A}:\text{B})=2E_W(\text{A}:\text{B})$. The EWCS is set to zero when the entanglement
wedge disconnects and therefore automatically shares the connectivity
transition found above. It would nevertheless be interesting to determine how the limiting
value of the EWCS on the connected branch depends on $\log g$ at the
Silver Blaze configuration.

Throughout this work we use the standard RT prescription for holographic ICFT backgrounds. We do not attempt an independent derivation from the gravitational replica path integral. In ordinary holographic CFTs, the Lewkowycz--Maldacena argument gives a semiclassical replica derivation of the RT area term, assuming the relevant replica-symmetric saddles exist and can be analytically continued \cite{Lewkowycz:2013nqa}. For disconnected regions this point is already nontrivial: Lewkowycz and Maldacena note that, for more complicated regions such as two intervals in a two-dimensional CFT, the replica construction can involve \(n\)-dependent topology. In the semiclassical limit, the connected/disconnected RT transition can therefore be understood as a dominance transition between different gravitational saddles. For smooth Janus backgrounds, it is natural to expect the same logic to apply, with the interface implemented through the scalar sources and bulk profiles. For thin-wall models, a fully explicit replica derivation would additionally have to include the brane action and the Israel junction conditions in the replicated geometry. We leave such a derivation to future work and take RT as the standard holographic prescription for the ICFT backgrounds studied here.\footnote{We thank Misha Smolkin for raising the question of a replica derivation of the RT prescription in this setting during Rotem Berman's presentation of this project at the Israel Physical Society meeting.}

Another promising direction is to study the entanglement transition from a purely field-theoretic perspective. In generic CFTs, the entanglement between two intervals is controlled by a twist-field four-point function and varies smoothly with the cross-ratio \cite{Calabrese:2009ez}. In the semiclassical limit of the gravity dual, the sharp transition is captured by the competition between RT saddles \cite{Headrick:2010zt}, and in two-dimensional examples it is related to the large-\(c\), sparse-spectrum structure of the corresponding twist correlators \cite{Hartman:2013mia,Tsujimura:2022siq}. Related large-\(c\) BCFT transitions have also been studied in terms of competing boundary/bulk OPE channels \cite{Sully:2020pza}. It would be interesting to develop the analogous field-theoretic description for ICFTs, in terms of interface twist correlators or defect conformal blocks, and to understand whether the Silver Blaze cancellation has a direct interpretation in that language.

A separate field-theoretic direction is to study free or weakly coupled interface theories. Interfaces realized as permeable conformal walls
have been extensively studied in free theories
\cite{Bachas:2001vj,Sakai:2008tt,Brehm:2015lja}, and free-field methods provide
powerful tools for computing entanglement \cite{Casini:2009sr}. 
In specific solvable families, these works show how
entanglement through a single interface depends on the transmission 
properties of the defect, although more generally the
entanglement can also depend on additional interface data. 
A natural extension, closer to the
setup studied here, would be to consider two disjoint intervals in such solvable
ICFTs and compute the smooth mutual information as a function of the separation and interface properties. In particular, one can
ask 
how the smooth two-interval mutual information depends on the transmission and other properties of the interface, whether the redistribution of correlations found 
holographically has a free ICFT analogue, and 
whether the interface-dependent part of the mutual information exhibits an
analogue of the Silver Blaze cancellation, even though the finite-\(c\) behavior
is expected to be smooth rather than exhibit a sharp phase transition.

A final direction is to connect these interface entanglement transitions to the
island picture of black-hole evaporation. The island prescription and its replica-wormhole derivation provide a mechanism
by which Page transitions are captured by new quantum extremal saddles
\cite{Penington:2019npb,Almheiri:2019psf,
Penington:2019kki,Almheiri:2019qdq}. In double-holographic brane-world models,
this mechanism admits a simple geometric realization. The same setup can be viewed in two equivalent ways. From
the brane point of view, one has a gravitating theory on the brane, where the
black hole lives, coupled to non-gravitating bath regions. From the full bulk
point of view, the entropy is computed by ordinary RT surfaces. When these RT
surfaces end on an end-of-the-world brane or intersect a Randall--Sundrum brane, they are
interpreted in the brane description as island saddles
\cite{Almheiri:2019hni,Chen:2020uac,Chen:2020hmv}.
A closely related brane-plus-bath perspective appears in the ICFT
construction of~\cite{Anous_2022}, where geodesics anchored at
two boundary points on the same side can cross the single brane twice or more
in an appropriate unequal-radius regime. Motivated by this connection,
it would be interesting to extend our two-interval analysis beyond the equal-radius thin-wall
models studied here and include such same-side geodesics that cross one or more branes before returning to the original side. 
This could clarify how
two-interval observables probe the distribution of information between different
bath subregions and the gravitating brane.

 \subsection*{Acknowledgments}

We thank Misha Smolkin for useful comments. S.C. and C.H. would like to thank the organizers and participants of the ``Iberian Strings 2026'' conference in Santiago de Compostela for the stimulating atmosphere in which this collaboration was initiated.  I.C.B. would like
to thank IPhT Saclay for the kind hospitality while part of this work was being carried out.
I.C.B. would also like to thank the COST Action CA22113 “Fundamental challenges in
theoretical physics” for financial support during part of this work through an STSM Grant.
 
 The work of E.A. is supported by the Severo Ochoa fellowship PA-23-BP22-170 and the work of I.C.B. is supported by the Severo Ochoa fellowship  NAC-AT-PUB-ASV-2025 BP24-116. The work of E.A., I.C.B. and C.H. is partially supported by the Spanish Agencia Estatal de Investigaci\'on and Ministerio de Ciencia, Innovaci\-on y Universidades through the grants PID2021-123021NB-I00 and PID2024-161500NB-I00. 
 The work of R.B., S.C. and O.S. is supported by the Israel Science Foundation (grant No. 1417/21), by the German Research
Foundation through a German-Israeli Project Cooperation (DIP) grant ``Holography and
the Swampland'', by Carole and Marcus Weinstein through the BGU Presidential Faculty
Recruitment Fund, by the ISF Center of Excellence for theoretical high-energy physics, by the VATAT Research Hub in the Field of quantum computing and by the ERC Starting Grant dSHologQI (project number 101117338). The work of R.B. and O.S. is further supported by the Kreitman fellowship
program at Ben-Gurion University.

%%%%%%%%%%%%%%%%%%%%%%%%%%%%

\appendix

%%%%%%%%%%%%%%%%%%%%%%%%%%%%%%%%%%%%%%%%%%%%%%%%
\section{Geodesic computation}\label{sec::appendixGeodComp}
%%%%%%%%%%%%%%%%%%%%%%%%%%%%%%%%%%%%%%%%%%%%%%%%
In this appendix we provide the details of the minimal length geodesic computation presented in Section \ref{sec:GeodConfInterEntr}. We follow closely the conventions of \cite{Afxonidis:2025jph}.

\subsection{Minimal length geodesics}

For the most general conformal defect spacetime in $2+1$ dimensions, given by the metric \eqref{eq::defectmetric}, the geodesic can be described by a profile $x(r)$, with boundary conditions fixed by the boundary interval considered. This profile is determined by extremizing the functional 
\begin{equation}\label{eq:SEEapp}
    S_{\rm A}=\frac{1}{4G}\int_{r_L}^{r_R} dr \, \mathcal{L} \ ,
\end{equation}
with the Lagrangian defined as
\begin{equation}\label{eq::Langrangianapp}
\mathcal{L}=R\, \sqrt{1+e^{2A(r)}\left( \frac{x'}{x}\right)^2} \ .
\end{equation}
On shell, the functional $S_{\rm A}$ evaluates the entanglement entropy of the corresponding interval A. Since geodesics anchored at the asymptotic boundary have infinite length, the result must be regularized. Throughout this work we implement this regularization by introducing cutoffs at some values $r=r_R$ and $r=r_L$, which we discuss in detail later around equation \eqref{eq:cutoffLapp}.

The geodesic equation can be reduced to a first order equation upon employing the isometry $x\to \lambda x$ within fixed time slices. 
The Lagrangian \eqref{eq::Langrangianapp} is invariant under this transformation, leading to the conserved Noether charge\footnote{From now on we will write $A$ rather than $A(r)$ and leave the $r$ dependence implicit.}
\begin{equation}\label{eq::cseqapp}
    c_s= \frac{\partial \mathcal{L}}{\partial x'}\frac{\partial (\lambda x)}{\partial \lambda}=\frac{R \,e^{2A}x'}{\sqrt{x^2+e^{2A}(x')^2}} \ .
\end{equation}
Solving this relation for $x'$ yields \eqref{eq::diffeq}. When $c_s < R e^{A_*}$, the argument of the square root in \eqref{eq::diffeq} is positive everywhere and the radial profile does not develop a turning point. On the other hand, for $c_s > R e^{A_*}$, the square root vanishes at the radial position $r=r_{\text{turn}}$. The limiting
case $c_s=Re^{A_*}$ corresponds to an interval with one endpoint at
the interface. For $c_s>Re^{A_*}$ the complete geodesic is then constructed by joining the two branches at this turning point, so that it runs from the asymptotic boundary down to $r_{\text{turn}}$ and back. This leads to the different cases discussed in Section~\ref{sec:GeodConfInterEntr}.

The conserved scaling charge $c_s$ is fixed by the endpoint data, and in particular by the only available cross ratio $l_L/l_R$. Its value therefore controls how the interval is located with respect to the interface and consequently affects the entanglement entropy. This ratio is obtained by integrating equation \eqref{eq::diffeq}. 
For crossing geodesics without a turning point, the branch with a fixed plus or minus sign gives
\begin{equation}\label{eq::logratioequationapp}
    \log\left( \frac{l_L}{l_R}\right)=\mp\int_{-\infty}^{\infty}dr\,\frac{c_s e^{_-A}}{\sqrt{e^{2A}R^2-c_s^2}}\ . 
\end{equation}
This equation makes manifest that $c_s$
 depends non-trivially on the ratio $l_L/l_R$, which in turn implies that the geodesic shape and therefore the entanglement entropy are sensitive to the positions of the interval endpoints with respect to the interface rather than only the total length of the interval. In addition, the ICFT theory has scale invariance, therefore we expect physical (cutoff independent) data to be invariant under a mutual rescaling of $l_L$ and $l_R$. This is hinted by the fact that such a rescaling does not change $c_s$ in the equation above.\footnote{For the double thin-wall model, the crossing-geodesic integral
in~\eqref{eq::logratioequationapp} must be split at the two branes and
evaluated using the appropriate piece of the warp factor in each
region. The resulting endpoint-ratio expression is
\eqref{eq::genericlogrho}.}

For geodesics with a turning point at $r=r_{\text{turn}}$, the solution is the union of the two branches shown in \eqref{eq::diffeq}, they both join at the point where the argument of the square root vanishes and $x'\to \infty$. 
The two branches contribute equally to   \eqref{eq::logratioequationapp}, in such a way that
\begin{equation}\label{eq::logratioequation2app}
    \log\left( \frac{l_L}{l_R}\right)=-2\int_{r_{\text{turn}}}^\infty\frac{c_s e^{_-A}}{\sqrt{e^{2A}R^2-c_s^2}}dr\,, \quad  \log\left( \frac{l_L}{l_R}\right)=2\int_{-\infty}^{r_{\text{turn}}}\frac{c_s e^{_-A}}{\sqrt{e^{2A}R^2-c_s^2}}dr\ .
\end{equation}

\subsection{Entanglement entropy}

We proceed with the computation of the entanglement entropy for crossing and non-crossing intervals. The entanglement entropy associated with a given interval is found by evaluating the length functional on the corresponding geodesic solution and integrating over the radial coordinate. On shell, the Lagrangian reduces to 
\begin{equation}\label{eq::onshellLangapp}
    \mathcal{L}=\frac{R}{\sqrt{1-c_s^2 e^{-2A} R^{-2}}} \ .
\end{equation}
Near $r\to \pm \infty$, the warp factor approaches its AdS form
\begin{equation}\label{eq:warpApprapp}
    e^A\approx \frac{1}{2} e^{a_\pm}e^{\pm r} \to +\infty \ ,
\end{equation}
with $a_\pm$ constants.\footnote{This equation fixes a small factor typo in equation (3.8) of (the first and second versions of) \cite{Afxonidis:2025jph}.}
It follows that the on-shell Lagrangian approaches $\mathcal{L}\approx R$ and the geodesic length diverges as the integration limits in \eqref{eq:SEEapp} are taken to the boundary. We regulate this divergence by introducing radial cutoffs at $r=r_L$ and $r=r_R$.  Using the Brown-Henneaux formula \eqref{eq:centralc}, the regularized entanglement entropy is
\begin{equation}\label{eq:SAcapp}
    S_\text{A}=\frac{c}{6}\int_{\gamma_{\rm reg}} \frac{dr}{\sqrt{1-c_s^2 e^{-2A} R^{-2}}} \ .
\end{equation}
The regularized integral depends on the type of interval
\begin{itemize}
    \item Crossing:
    \begin{equation}
        \int_{\gamma_{\rm reg}}=\int_{r_L}^{r_R},\quad \int_\gamma=\int_{-\infty}^\infty\ .
    \end{equation}
    \item Non-crossing to the right of the interface:
    \begin{equation}
        \int_{\gamma_{\rm reg}}=\int_{r_{\text{turn}}}^{r_L}+\int_{r_{\text{turn}}}^{r_R},\quad \int_\gamma=2\int_{r_{\text{turn}}}^\infty\ .
    \end{equation}
    \item Non-crossing to the left of the interface:
    \begin{equation}
        \int_{\gamma_{\rm reg}}=\int_{r_L}^{r_{\text{turn}}}+\int_{r_R}^{r_{\text{turn}}},\quad \int_\gamma=2\int_{-\infty}^{r_{\text{turn}}}\ .
    \end{equation}
\end{itemize} 
Equivalently, one can separate the divergent and finite pieces explicitly. 
The entanglement entropy for both crossing and non-crossing intervals can be written as
\begin{equation}\label{eq:SAapp}
    S_\text{A}=\frac{c}{6}\left[\int_\gamma \left( \frac{1}{\sqrt{1-c_s^2 e^{-2A} R^{-2}}}-1 \right)+\Delta r\right] \ .
\end{equation}
Here $\Delta r=r_R-r_L$ for crossing intervals, $\Delta r=r_R+r_L-2 r_{\text{turn}}$ for non-crossing intervals to the right of the interface and $\Delta r=-r_R-r_L+2 r_{\text{turn}}$ for non-crossing intervals to the left of the interface.

It remains to relate the radial cutoff to the UV cutoff of the dual field theory. To illustrate the prescription we consider first a symmetric crossing interval of length $l$, for which $r_R=-r_L=r_c$ and $S_\text{A}=c\, r_c/3$. In the absence of an interface, the bulk geometry is pure AdS. Using a position independent cutoff prescription in which the boundary CFT metric is the standard Minkowski metric, the cutoff surface in the Poincaré coordinates \eqref{eq:poincarecoord} is placed at $z=\epsilon$. Near the right endpoint of the interval, $x=l/2$, we have
\begin{equation}
    \epsilon=\frac{l}{2\cosh r_c}\approx l e^{-r_c}.
\end{equation}
Solving for $r_c$ then gives the standard CFT result
\begin{equation}
    r_c=\log\frac{l}{\epsilon}\ \Rightarrow \ S_\text{A}^{  ^{\rm CFT}}=\frac{c}{3}\log\frac{l}{\epsilon}.
\end{equation}
Assuming $l_L\leq l_R$, for an otherwise general geometry and interval, the right endpoint cutoff is related to the radial cutoff by
\begin{equation}
    \epsilon= l_R e^{-A(r_R)} \approx 2l_R e^{-a_+}e^{- r_R}.
\end{equation}
Thus,
\begin{equation}\label{eq:cutoffRapp}
    r_R=-a_++\log\frac{2l_R}{\epsilon}.
\end{equation}
The corresponding expression for the left endpoint depends on whether the interval crosses the interface: 
\begin{equation}\label{eq:cutoffLapp}
    \text{crossing:}\ \ -r_L=-a_-+\log\frac{2l_L}{\epsilon},\quad \text{non-crossing:}\ \ r_L=-a_++\log\frac{2l_L}{\epsilon}.
\end{equation}
If the geometry is reflection symmetric, $a_+=a_-$. For intervals with $l_L\geq l_R$ one proceeds in the same way with the proper adjustments.

For a symmetric crossing interval in a general ICFT background, the entanglement entropy takes the form
\begin{equation}
    S_\text{A}^{ ^{\rm ICFT}}=\frac{c}{3}\log\frac{l}{\epsilon}-\frac{c}{6}(a_++a_-).
\end{equation}
The finite term accompanying the universal logarithm can be identified with the boundary entropy 
\begin{equation}\label{eq:gisymapp}
    \log g=-\frac{c}{6}(a_++a_-).
\end{equation}

\subsubsection{Interface entropy for crossing intervals}

We now evaluate the interface entropy for crossing intervals, whose endpoints are located at distances $l_L$ and $l_R$ from the interface. To streamline the discussion we take $l_L\leq l_R$. It is straightforward to generalize the formulas for $l_R\leq l_L$.

The entanglement entropy is obtained by \eqref{eq:SAapp}. 
Combining the cutoff relation \eqref{eq:cutoffRapp} and \eqref{eq:cutoffLapp}, the entropy of a crossing interval can be written as \cite{Afxonidis:2024gne,Kruthoff:2021vgv}
\begin{equation}\label{eq::EEiccapp}
    S_\text{A}=\frac{c}{6}\log{\left( \frac{2l_L }{\epsilon}\right)}+\frac{c}{6}\log{\left( \frac{2l_R }{\epsilon}\right)}+\log g_c^{(2)} \ ,
\end{equation}
where\footnote{For the double thin-wall model, the crossing-geodesic
entropy integral in~\eqref{eq::loggiccapp} must be split at the two
branes and evaluated using the appropriate piece of the warp factor
in each region. The resulting expression is
\eqref{eq::genericlogg2c}.}
\begin{equation} \label{eq::loggiccapp}
    \log g^{(2)}_c =\frac{c}{6}\int_{-\infty}^{\infty}dr\left(\frac{1}{\sqrt{1-c_s^2e^{-2A}R^{-2}}}-1 \right)+\log g\ .
\end{equation}

The quantity $\log g^{(2)}_c$ differs from the conventional interface entropy. The latter is defined by writing the entropy in the form 
\begin{equation}\label{eq::EEicc2app}
    S_\text{A}=\frac{c}{3}\log{\left( \frac{l_L+l_R }{\epsilon}\right)}+\log g_c^{(1)} \ .
\end{equation}
The two definitions are related by a finite term depending only on the endpoint ratio
\begin{equation}\label{eq::g1g2relationapp}
    \log g^{(1)}_c= \log g^{(2)}_c + \frac{c}{6}\log \frac{4l_L/l_R}{(1+l_L/l_R)^2} \ .
\end{equation}

\subsubsection{Interface entropy for non-crossing intervals}

For non-crossing intervals to the right of the interface, $l_R>l_L$, 
the entropy can again be written in the form \eqref{eq::EEiccapp}. The corresponding finite interface contribution is now
\begin{equation} \label{eq::loggicc2app}
    \log g^{(2)}_{nc} =\frac{c}{3}\left[\int_{r_{\text{turn}}}^{\infty}dr\left(\frac{1}{\sqrt{1-c_s^2e^{-2A}R^{-2}}}-1 \right)-r_{\text{turn}}\right]-\frac{ca_+}{3}\ .
\end{equation}
Equivalently, using the standard CFT expression for an interval of length $l_R-l_L$, one may write
\begin{equation}\label{eq::EEicc3app}
    S_\text{A}=\frac{c}{3}\log{\left( \frac{l_R-l_L }{\epsilon}\right)}+\log g_{nc}^{(1)} \ .
\end{equation}
The two definitions of the interface entropy are then related by
\begin{equation}\label{eq::g1g2relationniccapp}
    \log g_{nc}^{(1)} = \log g^{(2)}_{nc} + \frac{c}{6}\log \frac{4l_L/l_R}{(1-l_L/l_R)^2} \ .
\end{equation}

For non-crossing intervals to the left of the interface, with $l_R<l_L$, 
the entropy can again be written in the form \eqref{eq::EEiccapp}. The corresponding finite interface contribution is now
\begin{equation} \label{eq::loggicc2app1}
    \log g^{(2)}_{nc} =\frac{c}{3}\left[\int_{-\infty}^{r_{\text{turn}}}dr\left(\frac{1}{\sqrt{1-c_s^2e^{-2A}R^{-2}}}-1 \right)+r_{\text{turn}}\right]-\frac{ca_-}{3}\ .
\end{equation}

%%%%%%%%%%%%%%%%%%%%%%%%%%%%%%%%%%%%%%%%%
%%%%%%%%%%%%%%%%%%%%%%%%%%%%%%%%%%%%%%%%%
%%%%%%%%%%%%%%%%%%%%%%%%%%%%%%%%%%%%%%%%%
\section{A geometric perspective on entanglement in the thin brane model}\label{sec:AppendixGeometricThinBrane}
%%%%%%%%%%%%%%%%%%%%%%%%%%%%%%%%%%%%%%%%%
%%%%%%%%%%%%%%%%%%%%%%%%%%%%%%%%%%%%%%%%%

In this appendix we present an alternative derivation of the entanglement entropy in the thin-brane model. This derivation gives the same result as the method used in the main text, but it makes the geometry of the RT surface more transparent. For this reason, we include it as a useful complementary perspective. We first introduce the angular coordinates in which the one-brane geometry is described
as two AdS patches glued along a ray in the Poincar\'e half-plane.
In Section~\ref{sec:AppendixGeometricThinBraneSingle},
we then show that in these coordinates,
constant-time geodesics are arcs of circles, which gives a simple geometric derivation of
the single-interval entropy. In Section~\ref{sec:AppendixGeometricThinBraneTwoIntervals},
we apply the same picture to two disjoint intervals and to the competing connected and
disconnected RT configurations. This provides a geometric derivation of the corresponding
mutual information and of the special cancellation underlying the Silver Blaze
configuration. Finally, in Section~\ref{sec:AppendixGeometricThinBraneTwoBranes}, we
explain how the construction generalizes to the case of two thin branes.

The action for the one-brane thin-wall model was given in
\eqref{eq:thinbranesymaction}. We denote the brane tension by $\Sigma$, and
the AdS radius on the two sides of the interface by $R$. 
On each side of the
brane the geometry is locally AdS$_3$, and in the coordinates used in the main 
text $r,x,t$, it is written in an AdS$_2$ slicing  \eqref{eq::defectmetric} with $e^{A(r)}=\cosh r$. For the geometric discussion in this appendix, it is useful to replace \(r\) by
an angular coordinate \(\theta\):
\begin{equation}\label{eq:Bcoordtrans}
r=\tanh^{-1}(\cos\theta). 
\end{equation}
The metric then becomes
\begin{equation}
ds^2=\frac{R^2}{\sin^2\theta}
\left[
d\theta^2+\frac{-dt^2+dx^2}{x^2}
\right]
=
\frac{R^2}{x^2\sin^2\theta}
\left(-dt^2+dx^2+x^2d\theta^2\right).
\end{equation}
Introducing Poincaré coordinates \eqref{eq:poincarecoord}, this becomes the standard AdS\(_3\) metric \eqref{eq::Poincarepatch}:
\begin{equation}\label{eq:poincordyappen}
y=x\cos\theta,\qquad z=x\sin\theta,\qquad ds^2=\frac{R^2}{z^2}\left(-dt^2+dy^2+dz^2\right).
\end{equation}
Thus $\theta$ is the polar angle in the Poincaré $(y,z)$ half-plane, while
$x$ is the radial coordinate in that plane. The boundaries of empty AdS$_3$ are located at $\theta=0,\pi$. Note that near the boundary $y=\pm x$ and so we will keep labeling endpoint coordinates for our intervals as (signed) $x_i$ in relevant places.

In the above coordinates, a brane at fixed $r$ is equivalently a ray at fixed
$\theta$. The two locally AdS regions are therefore represented by angular
wedges in the Poincaré half-plane. The opening angle \(\alpha\) of the wedge is
fixed by the brane tension through
\begin{equation}\label{eq:alphaopening}
\alpha=\frac{\pi}{2}+\sin^{-1}\left(\frac{R\Sigma}{2}\right).
\end{equation}

\subsection{Entanglement for a single interval with one thin-brane}\label{sec:AppendixGeometricThinBraneSingle}

The advantage of the above coordinates is that fixed-time geodesics obey simple circle equations in the $(y,z)$ plane. 
To illustrate this,
let us recall the standard RT surface for a single interval of length $l$ in
empty AdS$_3$. The interval endpoints are located at $y=-l/2$ and
$y=l/2$ on the asymptotic boundary. The corresponding geodesic is the
semicircle:
\begin{equation}
(y,z)=
\frac{l}{2}
\left(
\cos\xi,\sin\xi
\right),
\qquad
\xi_\epsilon
\leq
\xi
\leq
\pi-\xi_\epsilon ,
\end{equation}
where the cutoff \(z=\epsilon\) fixes
$\xi_\epsilon=\sin^{-1}(2\epsilon/l)\simeq 2\epsilon/l$. 
Equivalently, the curve obeys the circle equation
\begin{equation}
y^2+z^2=\left(\frac{l}{2}\right)^2 .
\end{equation} 
The regularized length of this geodesic and the associated entanglement entropy read 
\beq\label{Length_gamma}
\text{Length}(\gamma_\text{A})=2R\int_{\xi_\epsilon}^{\pi/2}\frac{d\xi}{\sin \xi} =2R\log\frac{l}{\epsilon}, \qquad
S_\text{A}=\frac{\text{Length}(\gamma_\text{A})}{4G}=\frac{c}{3}\log \frac{l}{\epsilon},
\eeq
where in the last equality we used the Brown-Henneaux relation
\eqref{eq:centralc} \cite{Brown:1986nw}.

The thin-brane RT surface is a direct generalization of the elementary
semicircle discussed above. Since each side of the brane is locally empty
AdS\(_3\), a non-crossing interval whose geodesic remains entirely within one
AdS patch gives the same entanglement entropy as in empty AdS.  
We therefore
focus on the new feature of the thin-brane geometry: an interval crossing the
defect.

\begin{figure}[htbp]
  \centering \includegraphics[width=0.55\textwidth]{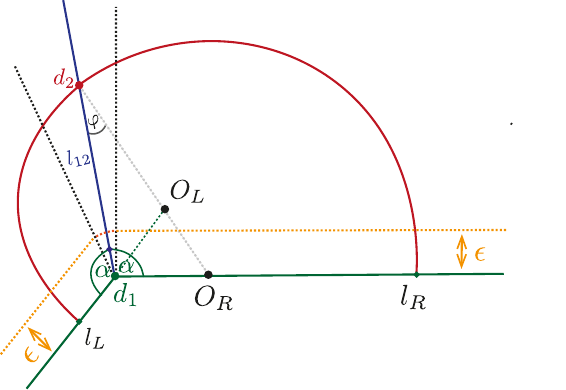} 
  \caption{
  Fixed-time slice of the two locally AdS\(_3\) patches glued across the thin
brane. The straight blue line represents the brane, and the equal opening angles of
the right and left patches are denoted by $\alpha$.
The dotted orange curves indicate the cutoff surfaces $z=z'=\epsilon$. The red curve is the RT surface for an interval with endpoints
at $y=l_R$ on the right boundary and $y'=-l_L$ on the left boundary. It is
made of two circular arcs, one in each patch, glued smoothly at the brane at the meeting point $d_2$. The
dashed gray line starting at point $d_2$ is the common normal at the gluing point and therefore passes
through the centers $O_R$ and $O_L$ of the two circles.}
  \label{fig:consttime}
\end{figure}

We take the interval to have endpoints at distance \(l_R\) and \(l_L\) from the
interface, with \(l_R>l_L>0\), without loss of generality. Let us denote the Poincar\'e coordinates in the right
patch by \((y,z)\) and those in the left patch by \((y',z')\). In terms of these coordinates, 
the right endpoint is located at \(y=l_R\), while the left endpoint is located at \(y'=-l_L\). 
We regulate the two asymptotic regions by placing the
cutoff surfaces at \(z=z'=\epsilon\). The fixed-time geometry is illustrated in
Fig.~\ref{fig:consttime}.

Instead of a single semicircle, the RT surface consists of two circular arcs,
one in each AdS patch. Each arc obeys a simple circle equation in the
corresponding Poincar\'e coordinates, namely
\begin{equation}
  (y-O_R)^2+z^2=D_R^2
  \qquad
  (y'-O_L)^2+z'^2= D_L^2 .
\end{equation}
Here \(O_R\) and \(O_L\) are the centers of the two circular arcs, while
\(D_R\) and \(D_L\) are their radii in the right and left
patches, respectively. The subscripts label the AdS patch containing the corresponding arc, not the
position of the center in the auxiliary Poincar\'e plane. In particular, a circle
center may lie outside the physical wedge of that patch. The endpoint conditions fix one point on each circle.
The remaining data are fixed by continuity at the point at which the two arcs meet the brane and by the smoothness condition there.

The two arcs must be glued smoothly across the brane: the full RT surface is
continuous, and its tangent is continuous at the gluing point. Equivalently, the
normal directions to the two circular arcs at the gluing point coincide. Since
the normal to a circle passes through its center, the common normal line at the
brane must pass through both circle centers, as shown in
Fig.~\ref{fig:consttime}. Thus the geometric problem reduces to determining the
two circle centers, the two radii, and the brane intersection point subject to
this collinearity condition. The parameters of the resulting arcs were computed
explicitly in \cite{Chapman:2018bqj}, see also \cite{Czech:2016nxc}. Below we
streamline the derivation, emphasizing the geometric steps that lead to the
same entanglement entropy formula obtained in the main text.

The remaining geometric data can be fixed by elementary trigonometry. Let \(d_1\) and \(d_2\) denote the two points
shown in Fig.~\ref{fig:consttime}, and let
$l_{12}\equiv |d_1d_2|$
be the length of the segment connecting them. We also denote by $\varphi$ the
angle $\measuredangle d_1d_2O_R$. 
The smoothness condition discussed above implies that \(d_2,O_R,O_L\) lie on a
single line, which is normal to both circular arcs at the gluing point. Applying
the sine rule to the two triangles $d_1 d_2 O_R$ and $d_1 d_2 O_L$ gives
\begin{equation} \label{solvingradii}
\begin{split}
    \frac{l_{12}}{\sin(\alpha+\varphi)}
    =
    \frac{D_R}{\sin\alpha}
    =
    \frac{l_R-D_R}{\sin\varphi},
    \qquad
    \frac{l_{12}}{\sin(\alpha-\varphi)}
    =
    \frac{D_L}{\sin\alpha}
    =
    \frac{D_L-l_L}{\sin\varphi}.
\end{split}
\end{equation}
We can see from the above equation that $\sin \alpha>\sin \phi$. Using the ratio $\rho\equiv {l_L}/{l_R}$ from equation \eqref{eq:ratiolllr}, these can be manipulated to give the following equation for $\varphi$
\begin{equation}\label{thetasol}
\rho
=
\left(
\frac{
\tan\frac{\alpha}{2}\tan\frac{\varphi}{2}-1
}{
\tan\frac{\alpha}{2}\tan\frac{\varphi}{2}+1
}
\right)^2 
\quad 
\Rightarrow
\quad
\tan\frac{\varphi}{2}
=
\left(
\frac{1-\sqrt{\rho}}
{1+\sqrt{\rho}}
\right)
\cot\frac{\alpha}{2} .
\end{equation}
More explicitly, using the opening angle  \eqref{eq:alphaopening} and the ratio \eqref{eq:ratiolllr} we can write
\begin{equation}\label{thetasol_sigma}
\tan\frac{\varphi}{2}
=
\frac{\sqrt{l_R}-\sqrt{l_L}}
{\sqrt{l_R}+\sqrt{l_L}}
\sqrt{
\frac{1-\Sigma R/2}{1+\Sigma R/2}
}.
\end{equation}
The expression is valid for the assumed range $l_R>l_L$ ($0<\rho<1$). 
For the opposite endpoint ordering, $l_L>l_R$ ($\rho>1$), one can either allow
$\varphi$ to change sign or use \eqref{thetasol} with \(\rho\to1/\rho\). 
We can also solve  \eqref{solvingradii} for the radii of the two circular arcs,
\begin{equation}
D_R
=
\frac{
\sqrt{l_R}\left(\sqrt{l_Rl_L}\,\Sigma R+l_R+l_L\right)
}{
2\sqrt{l_R}+\sqrt{l_L}\,\Sigma R
},
\qquad
D_L
=
\frac{
\sqrt{l_L}\left(\sqrt{l_Rl_L}\,\Sigma R+l_R+l_L\right)
}{
2\sqrt{l_L}+\sqrt{l_R}\,\Sigma R
}.
\end{equation}
Finally, the length $l_{12}$ takes
the simple form
\begin{equation}
l_{12}=\sqrt{l_Rl_L}.
\end{equation}

To compute the entanglement entropy, we sum the regularized lengths of the two
circular arcs. We denote by
\begin{equation}
\omega_R=\alpha+\varphi,\qquad
\omega_L=\alpha-\varphi
\end{equation}
the angular opening of the right and left arcs with respect to their respective circle centers, namely $\omega_R\equiv\measuredangle l_R O_R d_2$ and $\omega_L\equiv\measuredangle l_L O_L d_2$ in Fig.~\ref{fig:consttime}. 
The angular opening angles $\omega_{R,L}$ can also be expressed directly as:
\begin{equation}
\tan(\omega_R/2)=
\frac{2 \sqrt{l_R}+\sqrt{l_L} \Sigma R}{\sqrt{l_L} \sqrt{4-(\Sigma R)^2}}\, ,
\qquad
\tan(\omega_L/2)=
\frac{2 \sqrt{l_L}+\sqrt{l_R} \Sigma R}{\sqrt{l_R} \sqrt{4-(\Sigma R)^2}}\, .
\end{equation}
Parametrizing
each arc by the angle \(\xi\) around its own circle center, the integration 
takes the same form as in \eqref{Length_gamma}. 
Therefore, 
\begin{equation}\label{Length_gamma_defect_R}
\operatorname{Length}(\gamma_R)
=
R\int_{\xi_{R,\epsilon}}^{\omega_R}
\frac{d\xi}{\sin\xi}
\simeq
R\log\left[
\frac{2D_R}{\epsilon}\tan\frac{\omega_R}{2}
\right],
\end{equation}
and
\begin{equation}\label{Length_gamma_defect_L}
\operatorname{Length}(\gamma_L)
=
R\int_{\xi_{L,\epsilon}}^{\omega_L}
\frac{d\xi}{\sin\xi}
\simeq
R\log\left[
\frac{2D_L}{\epsilon}\tan\frac{\omega_L}{2}
\right]\, ,
\end{equation}
where we have used 
$\xi_{R,\epsilon}\simeq \frac{\epsilon}{D_R}$, and 
$\xi_{L,\epsilon}\simeq \frac{\epsilon}{D_L}$, which follows from $\epsilon\ll D_R,D_L$. 
The entanglement of the entire segment then reads, using  $c=3R/2G$:\footnote{The final result for the entanglement can be compared with the large boundary limit of the expression for global AdS from \cite{Chapman:2018bqj}, see in particular eqs.~(4.19)-(4.21) which read (after some substitutions):
\begin{equation}
    S_A = \frac{R}{2G}\log\left[\frac{2L_\mathcal{B}}{\epsilon}\sin\left(\frac{\theta_L-\theta_R}{2}\right)\right]+
    \frac{R}{2G}
    \log\left[\frac{2}{\sqrt{4-(\Sigma R)^2}}+\frac{\frac{\Sigma R}{\sqrt{4-(\Sigma R)^2}}}{\frac{\sqrt{| \sin (\theta_L)| -\cos (\theta_L) \cot (\theta_R)+\csc (\theta_R)}}{\sqrt{2} \sqrt{| \sin (\theta_L)| }}}\right].
\end{equation}
To obtain the result for the large boundary limit we follow the instructions of appendix $C$ of \cite{Chapman:2018bqj} and expand in $\theta_R = l_R/L_\mathcal{B}\ll 1$
$\theta_L = -l_L/L_\mathcal{B}\ll 1$ with $L_\mathcal{B}$ the characteristic (finite) boundary size.
This leads precisely to the same expression as in \eqref{eq:entanglementappendixGeometricsing}.} 
\beq\label{eq:entanglementappendixGeometricsing}
S_\text{A}=\frac{R}{4G} \log\left[\frac{2D_R}{\epsilon}\tan{\frac{\omega_R}{2}}\right]+\frac{R}{4G} \log\left[\frac{2D_L}{\epsilon}\tan{\frac{\omega_L}{2}}\right]
=
\frac{c}{3}\log\left[\frac{2(l_R+l_L+\Sigma R \sqrt{l_R l_L})}{\epsilon \sqrt{4-(\Sigma  R)^2}}\right].
\eeq
Using the definition \eqref{eq:SA} for the finite part of the entanglement this yields: 
\begin{equation}\label{eq::logg2cGeomApp}
    \log g_c^{(2)}\equiv S_\text{A} - \frac{c}{6}\log\left(\frac{2l_R}{\epsilon}\right)-\frac{c}{6}\log\left(\frac{2l_L}{\epsilon}\right) = \frac{c}{3}\log\left[\frac{(l_R+l_L+\Sigma R \sqrt{l_R l_L})}{ \sqrt{l_R l_L} \sqrt{4-(\Sigma R) ^2}}\right].
\end{equation}
The above results for the entanglement are manifestly symmetric under exchanging $l_R$ and $l_L$,
and therefore extend beyond the ordering $l_R>l_L$ used in the derivation to
any $\rho=l_L/l_R>0$. In the limit $\Sigma\to0$, $\log g_c^{(2)}$   agrees with the empty-AdS result in
\eqref{eq::pureAdS}, as it should. For a symmetric crossing interval, $l_R=l_L$, we obtain the boundary entropy of the one-brane thin-wall model 
\eqref{eq::boundaryentropyRS}, as expected. 
Moreover, \eqref{eq::logg2cGeomApp} reduces to \eqref{eq::logg2RSc} after making the reparametrization
\begin{equation}
 \tanh r_*=\frac{\Sigma R}{2} \ .
\end{equation}

\subsection{Entanglement for two intervals with one thin-brane}\label{sec:AppendixGeometricThinBraneTwoIntervals}
We now turn to the case of two disjoint intervals. As in the discussion of
Section~\ref{sec:EntanglementTransition}, the entanglement entropy is determined
by comparing two competing RT configurations: a disconnected configuration and a
connected configuration. In the thin-brane geometry, each geodesic segment is
again described by circular arcs in the corresponding Poincar\'e patch, so the
geometric approach of the previous subsection can be applied to each
candidate configuration.

\begin{figure}[htbp]
\centering
\subfigure[Disconnected configuration\label{fig:minsurfdisca}]{\includegraphics[scale=0.3]{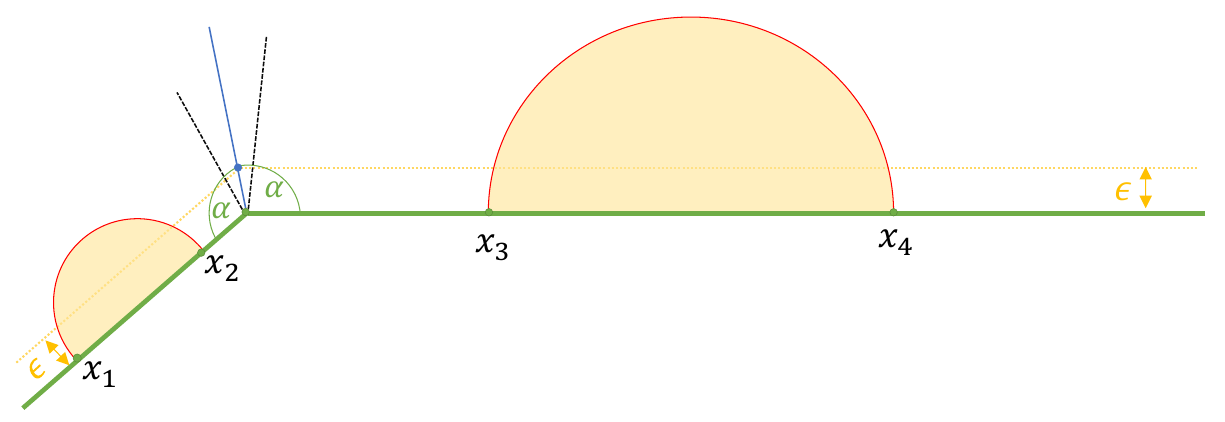}} \,
\subfigure[Connected configuration\label{fig:minsurfconb}]{\includegraphics[scale=0.3]
{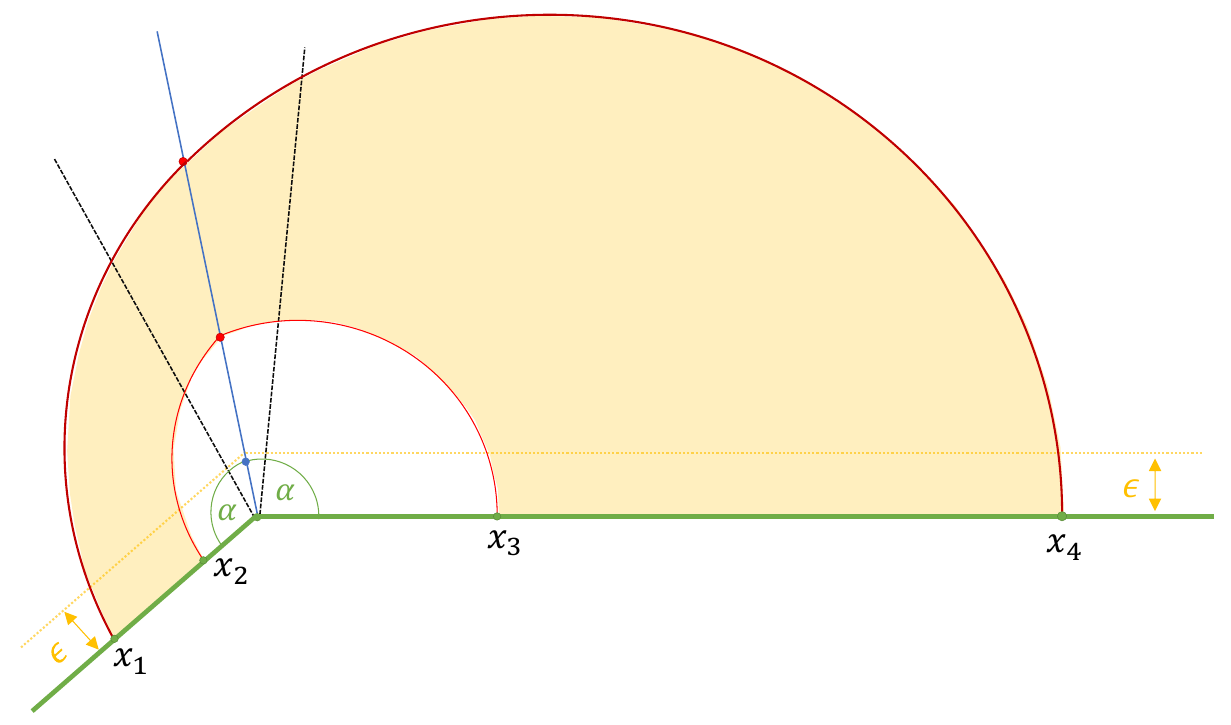}}
\caption{Candidate RT configurations for two boundary intervals in the thin-brane
geometry.  
In the illustration, the interface is located between the two intervals. (a) The minimal surface of the disconnected configuration. (b) The minimal surface of the connected configuration.} 
\label{fig:minimalsurfacedefect}
\end{figure}

An example is shown in Fig.~\ref{fig:minimalsurfacedefect}, with one
interval on each side of the interface. In the disconnected configuration,
shown in Fig.~\ref{fig:minsurfdisca}, each interval is homologous
to its own RT geodesic, so the total length is the sum of two single-interval
contributions. In the connected configuration, shown in
Fig.~\ref{fig:minsurfconb}, the geodesics instead connect
endpoints belonging to different intervals. The entanglement transition occurs
when the total lengths of these two configurations become equal.

\begin{figure}[htbp]
\centering
\subfigure[interface between intervals\label{fig:intermid}]{\includegraphics[scale=0.65]{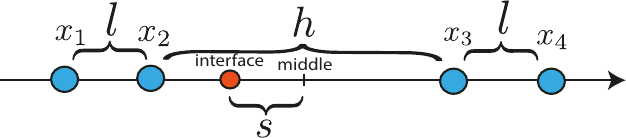}
}~~~~~~
\subfigure[interface inside left interval\label{fig:interleft}]{\includegraphics[scale=0.6]{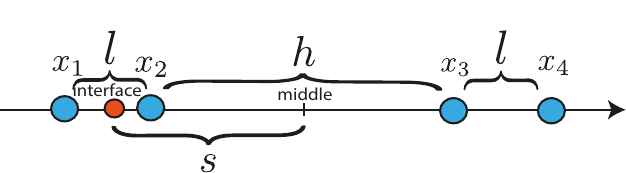}}
\caption{
Two relative placements of the interface for a pair of equal-length boundary
intervals. The interval endpoints, denoted by blue circles, are ordered as
\(x_1<x_2<x_3<x_4\), with
\(\text{A}=[x_1,x_2]\) and \(\text{B}=[x_3,x_4]\). The red circle denotes the
interface, placed at \(x=0\), and the braces indicate several relevant distances. 
Panel (a) shows the case in which the interface lies between the two intervals. Panel (b) shows the case in which the interface lies
inside one of the intervals, taken here to be A.} 
\label{fig:config2sub}
\end{figure}

More generally, we will study two possible relative positions of the interface,
shown in Fig.~\ref{fig:config2sub}. First, the interface may lie between the two intervals, as in Fig.~\ref{fig:intermid}. Second, the interface
may lie inside one of the intervals, as in Fig.~\ref{fig:interleft} (without
loss of generality, we take this to be the left interval).

In the rest of this subsection, we use signed boundary coordinates \(x_i\), as in
Section~\ref{sec:EntanglementTransition}. These should be identified with the
Poincar\'e boundary coordinate \(y\) of
\eqref{eq:poincarecoord} and \eqref{eq:poincordyappen}. The
AdS\(_2\)-slice coordinate \(x\) used in \eqref{eq::defectmetric} is positive
and approaches \(|y|\) on the asymptotic boundary: \(x\to y\) on the right
boundary and \(x\to -y\) on the left boundary.

When applying the single-interval formulas from
Section~\ref{sec:AppendixGeometricThinBraneSingle}, the quantities \(l_L\) and
\(l_R\) denote positive distances from the interface. Thus, for a crossing
interval with signed endpoints \(x_a<0<x_b\), we should substitute $l_L=|x_a|$, and $l_R=x_b$.
For a non-crossing interval, the entropy depends only on the length of the
interval within one AdS patch. 
The relevant formulas can therefore be written in
terms of signed endpoints as
\begin{equation}\label{eq:Scxaxb}
S_c(x_a,x_b)
=
\frac{c}{3}
\log\left[
\frac{
2\left(|x_a|+|x_b|+\Sigma R\sqrt{|x_a x_b|}\right)
}{
\epsilon\sqrt{4-(\Sigma R)^2}
}
\right],
\qquad x_a x_b<0 ,
\end{equation}
and
\begin{equation}\label{eq:Sncxaxb}
S_{nc}(x_a,x_b)
=
\frac{c}{3}
\log\frac{|x_b-x_a|}{\epsilon},
\qquad x_a x_b>0 .
\end{equation}
Here the subscript $c$ denotes a crossing interval, whose geodesic crosses the defect,
while the subscript $nc$ denotes a non-crossing interval, whose geodesic remains on one
side of the defect.

We now apply these two building blocks to study the entanglement transitions for the two relative positions of the
interface shown in Fig.~\ref{fig:config2sub}. In
Subsection~\ref{app:subappin}, we consider the case in
which the interface lies   between the two intervals. In
Subsection~\ref{app:subappin2}, we instead place the
interface inside one of the intervals. Finally, in
Subsection~\ref{app:subappSB}, we use the latter setup to derive the
special Silver Blaze configuration, for which the transition point becomes
independent of the brane tension.

\subsubsection{Interface between the two intervals}\label{app:subappin}

We first consider the case in which the interface lies  
between the two intervals,
\begin{equation}
x_1<x_2<0<x_3<x_4 .
\end{equation}
The disconnected RT configuration consists of two non-crossing geodesics,
one for each interval separately. The connected configuration consists of two
crossing geodesics, connecting \(x_1\) to \(x_4\) and \(x_2\) to \(x_3\).
Therefore the transition is determined by
\begin{equation}
S_c(x_1,x_4)+S_c(x_2,x_3)
=
S_{nc}(x_1,x_2)+S_{nc}(x_3,x_4).
\end{equation}
Using the signed-coordinate form of the single-interval results written above,
this becomes
\begin{equation}\label{eq:transitioncondapp}
4
\left(
|x_1|+x_4+\Sigma R\sqrt{|x_1|x_4}
\right)
\left(
|x_2|+x_3+\Sigma R\sqrt{|x_2|x_3}
\right)
=
\left(4-(\Sigma R)^2\right)
(x_2-x_1)(x_4-x_3).
\end{equation}
Here \(x_2-x_1\) and \(x_4-x_3\) are the lengths of the two original
intervals, while the absolute values convert the left endpoints into positive
distances from the interface when applying the crossing-interval formula.
Equivalently, the mutual information of the connected configuration is
\begin{equation}
I_{\text{conn}}(\text{A}:\text{B})
=
\frac{c}{3}
\log\left[
\frac{
\left(4-(\Sigma R)^2\right)(x_2-x_1)(x_4-x_3)
}{
4
\left(|x_1|+x_4+\Sigma R\sqrt{|x_1|x_4}\right)
\left(|x_2|+x_3+\Sigma R\sqrt{|x_2|x_3}\right)
}
\right].
\end{equation}
The transition occurs when \(I_{\text{conn}}(\text{A}:\text{B})=0\), which is equivalent to the condition \eqref{eq:transitioncondapp}.

For equal interval lengths, we use the same parametrization \eqref{eq:endpointsshl} as in the main text, 
\begin{equation}\label{eq:points_hsl_app}
x_1=s-\frac{h}{2}-l,\qquad
x_2=s-\frac{h}{2},\qquad
x_3=s+\frac{h}{2},\qquad
x_4=s+\frac{h}{2}+l .
\end{equation}
Introducing the dimensionless variables
\begin{equation}\label{eq:dimlessvarsapp}
\lambda\equiv\frac{h}{l},
\qquad
\sigma\equiv\frac{s}{l},
\end{equation}
we have
\begin{equation}\label{eq:appx1x4ofls}
\frac{x_1}{l}=\sigma-\frac{\lambda}{2}-1,\qquad
\frac{x_2}{l}=\sigma-\frac{\lambda}{2},\qquad
\frac{x_3}{l}=\sigma+\frac{\lambda}{2},\qquad
\frac{x_4}{l}=1+\sigma+\frac{\lambda}{2}.
\end{equation}
In the present case the interface lies between the intervals, so
\begin{equation}
-\frac{\lambda}{2}<\sigma<\frac{\lambda}{2}.
\end{equation}
Substituting these expressions into the transition condition gives
\begin{equation}
\left[
2+\lambda
+\Sigma R\sqrt{\left(1+\frac{\lambda}{2}\right)^2-\sigma^2}
\right]
\left[
\lambda
+\Sigma R\sqrt{\frac{\lambda^2}{4}-\sigma^2}
\right]
=
1-\frac{(\Sigma R)^2}{4}.
\end{equation}

\subsubsection{Interface inside one of the intervals}\label{app:subappin2}

We now consider the case in which the interface lies inside one of the two
intervals. Without loss of generality, we take this interval to be A, so that
the signed endpoints are ordered as
\begin{equation}
x_1<0<x_2<x_3<x_4 .
\end{equation}
The disconnected RT configuration consists of a crossing geodesic for
\([x_1,x_2]\) and a non-crossing geodesic for \([x_3,x_4]\). The connected
configuration instead consists of a crossing geodesic for \([x_1,x_4]\) and a
non-crossing geodesic for \([x_2,x_3]\). Therefore the transition is determined
by
\begin{equation}
S_c(x_1,x_2)+S_{nc}(x_3,x_4)
=
S_c(x_1,x_4)+S_{nc}(x_2,x_3).
\end{equation}
Using equations \eqref{eq:Scxaxb}-\eqref{eq:Sncxaxb}, this becomes
\begin{equation}\label{eq:transition_interB}
\left(
|x_1|+x_2+\Sigma R\sqrt{|x_1|x_2}
\right)(x_4-x_3)
=
\left(
|x_1|+x_4+\Sigma R\sqrt{|x_1|x_4}
\right)(x_3-x_2).
\end{equation}
Equivalently, the mutual information of the connected configuration is
\begin{equation}
I_{\text{conn}}(\text{A}:\text{B})
=
\frac{c}{3}
\log\left[
\frac{
\left(
|x_1|+x_2+\Sigma R\sqrt{|x_1|x_2}
\right)(x_4-x_3)
}{
\left(
|x_1|+x_4+\Sigma R\sqrt{|x_1|x_4}
\right)(x_3-x_2)
}
\right].
\end{equation}
The transition occurs when \(I_{\text{conn}}(\text{A}:\text{B})=0\), which is equivalent to
\eqref{eq:transition_interB}.

For equal interval lengths $l$, we again use the parametrization
\eqref{eq:points_hsl_app} and the dimensionless variables \eqref{eq:dimlessvarsapp}.
The case in which the interface lies inside the left interval corresponds to
\begin{equation}
x_1<0<x_2,
\qquad\text{or equivalently}\qquad
\frac{\lambda}{2}<\sigma<1+\frac{\lambda}{2}.
\end{equation}
Then the endpoint coordinates can be expressed in terms of the dimensionless variables as in \eqref{eq:appx1x4ofls}. 
Substituting those into
\eqref{eq:transition_interB}, we find the transition condition
\begin{equation}\label{eq:entgenericcasetwo}
1+\Sigma R
\sqrt{
\left(1+\frac{\lambda}{2}-\sigma\right)
\left(\sigma-\frac{\lambda}{2}\right)
}
=
\lambda
\left[
2+\lambda
+\Sigma R
\sqrt{
\left(1+\frac{\lambda}{2}-\sigma\right)
\left(1+\sigma+\frac{\lambda}{2}\right)
}
\right].
\end{equation}

\subsubsection{The universal Silver Blaze configuration}\label{app:subappSB}

We now derive the universal Silver
Blaze configuration discussed in Section~\ref{sec:MIholoICFTSB}. We focus on the
case in which the interface lies inside the interval A, where the transition condition is given in \eqref{eq:transition_interB}. Moving all
terms to one side, it can be written as
\begin{equation}
f(x_1,x_2,x_3,x_4)
+
\Sigma R\,g(x_1,x_2,x_3,x_4)
=0,
\end{equation}
where
\begin{equation}
\begin{split}
f
&=
\left(|x_1|+x_2\right)(x_4-x_3)
-
\left(|x_1|+x_4\right)(x_3-x_2),
\\
g
&=
\sqrt{|x_1|x_2}\,(x_4-x_3)
-
\sqrt{|x_1|x_4}\,(x_3-x_2).
\end{split}
\end{equation}
A transition point which is independent of the brane tension must therefore
satisfy
\begin{equation}
f=g=0.
\end{equation}
Solving the above conditions, we recover equation  \eqref{eq:silverblazepoint}, namely
\begin{equation}\label{eq:SBappendixB}
x_3=|x_1|=\sqrt{x_2x_4},
\end{equation}
which is precisely the Silver Blaze configuration. The observation $-x_1=x_3$ means that the twist operators involved in the calculation of the entanglement sit at mirror points  relative to the defect. The condition $|x_1|x_3=x_2 x_4$ is some sort of geometric average condition. 
Equivalently, these conditions imply
\begin{equation}
\frac{x_2}{|x_1|}
=
\frac{|x_1|}{x_4},
\qquad
\frac{x_3}{x_4}
=
\frac{x_2}{x_3}.
\end{equation}
Therefore, in the mutual information, the crossing contributions have identical
arguments and cancel each other, and the same is true for the non-crossing
contributions. This makes the transition independent not only of the thin-brane
tension, but of the interface data more generally. 
For equal interval lengths $l$, using equations \eqref{eq:points_hsl_app}-\eqref{eq:appx1x4ofls}, the condition \(-x_1=x_3\) gives
\begin{equation}
\sigma_{\rm SB}=\frac12.
\end{equation}
The second condition then gives
\begin{equation}
\lambda_{\rm SB}
=
\sqrt{2}-1,
\end{equation}
which is exactly the critical value of the CFT without an interface.
More generally, if the two interval lengths are $l_\text{A}$ and $l_\text{B}$, the analysis follows that in equations \eqref{eq::kappasetup}-\eqref{eq:SBgeneral2}.

It is important to emphasize that the universality concerns the location of the
transition, not the value of the entropy at the transition. Evaluating, for
example, the disconnected entropy at the Silver Blaze point gives
\begin{equation}
S^{\rm SB}
=
\frac{c}{3}
\log\left[
\frac{
2\sqrt{x_2x_4}\,
(\sqrt{x_4}-\sqrt{x_2})
\left(
\sqrt{x_4}+\sqrt{x_2}
+\Sigma R\,(x_2x_4)^{1/4}
\right)
}{
\epsilon^2\sqrt{4-(\Sigma R)^2}
}
\right].
\end{equation}
Thus the entropy itself still depends on the endpoint positions and on the brane
tension. To isolate the cutoff-independent finite part, we subtract the endpoint
logarithms in the same \(\log g^{(2)}\) convention used in the main text, see Eq.~\eqref{eq:SA}. We find
\begin{equation}
\left(\log g^{(2)}\right)_{\rm SB}
=
\frac{c}{3}
\log\left[
\frac{
(1-\nu_{\rm SB})
\left(
1+\nu_{\rm SB}
+\Sigma R\sqrt{\nu_{\rm SB}}
\right)
}{
2\nu_{\rm SB}\sqrt{4-(\Sigma R)^2}
}
\right]
, \qquad \nu_{\rm SB}\equiv \sqrt{\frac{x_2}{x_4}}.
\end{equation}
This finite part is independent of the overall scale of the endpoints, as expected from
scale invariance, but it still depends on the remaining endpoint ratio
\(\nu_{\rm SB}\) and on the brane tension. What is universal is that, when the endpoints
obey the Silver Blaze relations, the connected and disconnected configurations become
degenerate independently of the interface data.

\subsection{Entanglement for a single interval with two thin branes}
\label{sec:AppendixGeometricThinBraneTwoBranes}

\begin{figure}[htbp]
  \centering 
  \includegraphics[width=0.55\textwidth]{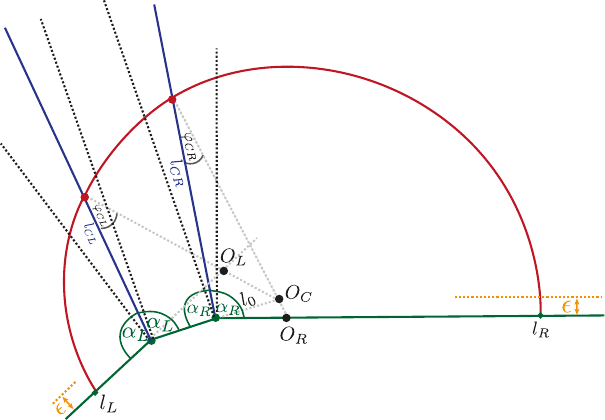}
\caption{A constant-time slice in the two-brane geometry, in which
three locally AdS$_3$ regions with common radius \(R\) are joined
across two branes. The two blue rays represent the branes separating
the left, central, and right AdS regions. The central boundary segment
is kept finite at first, so that the angular data can be defined. The
single-interface limit is then obtained by shrinking this segment to
zero after imposing the local matching conditions. The common opening angle across the left brane, with tension
\(\Sigma_1\), is denoted by \(\alpha_L\), while the common
opening angle across the right brane, with tension \(\Sigma_2\),
is denoted by \(\alpha_R\). The red curve is the RT surface for a crossing interval
with endpoint distances \(l_L\) and \(l_R\). It consists of three
circular arcs, with centers located at \(O_R\) for the right region,
\(O_L\) for the left region, and \(O_C\) for the central region, glued
smoothly at the two branes. The dashed gray lines are the common
normals at the gluing points and pass through the centers of the
relevant circles. For positive subcritical tensions,
\(\alpha_L,\alpha_R>\pi/2\). The figure is schematic: the orientation
of the dashed gray normals relative to the branes, and the positions
of the circle centers \(O_L\), \(O_C\), and \(O_R\) with respect to the
different AdS$_3$ patches, represent one possible geometric
arrangement.}
\label{fig:consttime2brane}
\end{figure}

A similar geometric construction can be applied to the two-brane thin-wall
geometry of Section~\ref{sec:Thinwall2}, see also \cite{Baig:2022cnb}. The bulk consists of three locally
AdS\(_3\) regions, $\mathcal M_L$, $\mathcal M_C$, and $\mathcal M_R$,
separated by two codimension-one branes \(\mathcal Q_1\) and
\(\mathcal Q_2\). We take all three regions to have equal AdS radius
$R$.
The corresponding
brane tensions are denoted by \(\Sigma_1\) and \(\Sigma_2\).
The relevant action is
\begin{equation}
S
=
\frac{1}{16\pi G}
\sum_{i=L,C,R}
\int_{\mathcal M_i} d^3x\,\sqrt{-g_i}
\left(
\mathcal R_i+\frac{2}{R^2}
\right)
-
\frac{1}{8\pi G}
\sum_{j=1}^{2}
\Sigma_j
\int_{\mathcal Q_j} d^2\xi\,\sqrt{-h_j}.
\end{equation}
We begin with a 
constant-time picture in which the central region has a finite boundary segment.
In this geometry the two branes end on two distinct boundary points,
see Fig.~\ref{fig:consttime2brane}. The exterior regions
are represented by angular wedges in their respective Poincar\'e half-planes,
while the central region is the angular domain bounded by the two branes and the
finite boundary segment. After imposing the local matching conditions at
the two branes, we take the limit in which the finite boundary segment shrinks to zero.
This gives the single-interface two-brane geometry used in the main text.

We denote by \(\alpha_L\) the common opening angle on the two
sides of the left brane \(\mathcal{Q}_1\), separating \(\mathcal M_L\)
from \(\mathcal M_C\), and by \(\alpha_R\) the common opening
angle on the two sides of the right brane \(\mathcal{Q}_2\), separating
\(\mathcal M_C\) from \(\mathcal M_R\), see
Fig.~\ref{fig:consttime2brane}. Since all three AdS regions
have the same radius \(R\), continuity of the induced metric
identifies the opening angles on the two sides of each brane. 
The local matching condition at the left brane \(\mathcal{Q}_1\) reads 
\begin{equation}
\frac{R}{\sin\alpha_L}
= 
-\frac{2\cot\alpha_L}{\Sigma_1}.
\end{equation}
Similarly, at the right brane \(\mathcal Q_2\), 
one has
\begin{equation}
\frac{R}{\sin\alpha_R}
=
-\frac{2\cot\alpha_R}{\Sigma_2}.
\end{equation}
Equivalently,
\begin{equation}
\alpha_L
=
\frac{\pi}{2}
+\sin^{-1}\!\left(\frac{R\Sigma_1}{2}\right),
\qquad
\alpha_R
=
\frac{\pi}{2}
+\sin^{-1}\!\left(\frac{R\Sigma_2}{2}\right).
\end{equation}
For tensions satisfying \eqref{eq:existence_cond}, both opening
angles are larger than \(\pi/2\). Consequently,
\(\alpha_L+\alpha_R>\pi\), and the central wedge remains
non-degenerate.
The two brane tensions are independent, so in general
\(\alpha_L\neq\alpha_R\). Equal tensions give the
reflection-symmetric subcase \(\alpha_L=\alpha_R\).

Since \(\alpha_L,\alpha_R>\pi/2\), a non-crossing
semicircular geodesic anchored on either exterior boundary does
not intersect the corresponding brane. It therefore remains
entirely within a single exterior AdS$_3$ patch, and its entropy
agrees with the empty-AdS result.

Next, consider a crossing interval with endpoint distances \(l_L\) and \(l_R\)
from the interface. The corresponding RT
surface consists of three circular arcs: one in the left exterior region, one in
the central region, and one in the right exterior region. We denote the circle
centers by \(O_L\), \(O_C\), and \(O_R\), respectively. Smoothness at each brane
requires the two arcs meeting there to have a common tangent. Equivalently, the
normal line at the gluing point passes through the centers of the two relevant
circles. Thus the problem again reduces to elementary circle geometry, now with
two gluing points instead of one.

The remaining geometric constraints are again fixed by simple application of the  sine theorem.
Let us illustrate the construction for the orientation shown in
Fig.~\ref{fig:consttime2brane}, where we take
\(\varphi_{CR},\varphi_{CL}>0\). Other orientations of the common normals lead to
the same type of equations, with the corresponding signs of the angles changed. We denote the radii of the right, central, and left circular arcs by
\(D_R,D_C,D_L\), respectively. The endpoint distances from the interface are
\(l_R\) and \(l_L\), and we denote by $l_{CR}$, $l_{CL}$, and
$l_0$ the different lengths indicated in 
Fig.~\ref{fig:consttime2brane}. Applying the sine theorem gives
\begin{equation}\label{eq:rel2brane}
\begin{split}
    &\frac{l_{CR}}{\sin(\alpha_R+\varphi_{CR})}
    =
    \frac{D_R}{\sin\alpha_R}
    =
    \frac{l_R-D_R}{\sin\varphi_{CR}},
    \\
    &\frac{l_{CL}}{\sin(\alpha_L-\varphi_{CL})}
    =
    \frac{D_L}{\sin\alpha_L}
    =
    \frac{D_L-l_L}{\sin\varphi_{CL}},
    \\
    &\frac{D_C}{\sin\alpha_R}
    =
    \frac{l_{CR}}{\sin(\alpha_R-\varphi_{CR})}
    =
    \frac{l_0}{\sin\varphi_{CR}},
    \\
    &\frac{D_C}{\sin\alpha_L}
    =
    \frac{l_{CL}}{\sin(\alpha_L+\varphi_{CL})},
    \\
    &\frac{l_0}{\sin(\alpha_L-\varphi_{CL})}
    =
    \frac{D_C-D_L}{\sin(2\alpha_L-\pi)} .
\end{split}
\end{equation}
The last relation uses the limit in which the central boundary segment is shrunk
to zero, so that the two branes end at a single boundary interface point.
For fixed endpoint distances \(l_L,l_R\) and fixed opening angles
\(\alpha_L,\alpha_R\), the unknown geometric data are
$D_R,\;D_L,\;D_C,\;l_{CR},\;l_{CL},\;l_0,\;
\varphi_{CR},\;\varphi_{CL}$,
and the relations above give 8 independent equations, so this
fixes the configuration completely.

The sine-rule system simplifies further in the equal-tension 
case $\Sigma_1=\Sigma_2$. 
Reflection symmetry of the background implies in this case
\(\alpha_L=\alpha_R\equiv\alpha\), while the geometric matching
conditions imply \(\varphi_{CR}=\varphi_{CL}\equiv\varphi\).  
In this special case, the geometric constraints reduce to a single equation
fixing \(\varphi\) in terms of the endpoint-distance ratio:
\begin{equation}
\frac{l_R}{l_L}
=
\left(
\frac{
1+\tan\frac{\alpha}{2}\tan\frac{\varphi}{2}
}{
1-\tan\frac{\alpha}{2}\tan\frac{\varphi}{2}
}
\right)^3
\left(
\frac{
\tan\frac{\alpha}{2}-\tan\frac{\varphi}{2}
}{
\tan\frac{\alpha}{2}+\tan\frac{\varphi}{2}
}
\right).
\end{equation}
As a simple consistency check, for \(l_R=l_L\) the equation is solved by
\(\varphi=0\), as expected from the left--right symmetry of the fully symmetric
configuration. We include the above simplified expression as a useful check of the geometric
construction.

Next, we return to the case of possibly unequal brane tensions
and evaluate the entanglement entropy for a crossing interval,
for the orientation shown in
 Fig.~\ref{fig:consttime2brane}. 
The crossing RT surface consists of three circular arcs, which we
denote by $\gamma_R$, $\gamma_C$, $\gamma_L$. The total entropy is
\begin{equation}
S_{\text{A}}
=
\frac{
\operatorname{Length}(\gamma_R)
+
\operatorname{Length}(\gamma_C)
+
\operatorname{Length}(\gamma_L)
}{4G}.
\end{equation}

Let
\begin{equation}
\omega_R=\alpha_R+\varphi_{CR},
\qquad
\omega_L=\alpha_L-\varphi_{CL}
\end{equation}
be the angular endpoints of the right and left exterior arcs at the branes. As in
the one-brane case, we parametrize each circular arc by an angular coordinate
\(\xi\) around its own circle center. The cutoff surfaces intersect the exterior
circles at small angles 
so that \(\xi_{R,\epsilon}\simeq \epsilon/D_R\) and
\(\xi_{L,\epsilon}\simeq \epsilon/D_L\). Therefore
\begin{equation}\label{eq:gammar}
\operatorname{Length}(\gamma_R)
=
R\int_{\xi_{R,\epsilon}}^{\omega_R}
\frac{d\xi}{\sin\xi}
\simeq
R\log\left[
\frac{2D_R}{\epsilon}
\tan\frac{\alpha_R+\varphi_{CR}}{2}
\right],
\end{equation}
and
\begin{equation}\label{eq:gammal}
\operatorname{Length}(\gamma_L)
=
R\int_{\xi_{L,\epsilon}}^{\omega_L}
\frac{d\xi}{\sin\xi}
\simeq
R\log\left[
\frac{2D_L}{\epsilon}
\tan\frac{\alpha_L-\varphi_{CL}}{2}
\right].
\end{equation}
The central arc has no UV-divergent endpoint contribution. 
For the orientation shown in
Fig.~\ref{fig:consttime2brane}, its angular range is $\xi \in [\pi-\alpha_L-\varphi_{CL},\alpha_R-\varphi_{CR}]$
and therefore
\begin{equation}
\begin{split}\label{eq:gammac}
\operatorname{Length}(\gamma_C)
&=
R
\int_{\pi-\alpha_L-\varphi_{CL}}^{\alpha_R-\varphi_{CR}}
\frac{d\xi}{\sin\xi}
=
R
\log\left[
\tan\frac{\alpha_R-\varphi_{CR}}{2}
\tan\frac{\alpha_L+\varphi_{CL}}{2}
\right].
\end{split}
\end{equation} 
Other orientations of the common normals lead to analogous
expressions with the corresponding angular limits.

\paragraph{Exchange of the two brane tensions.}

Exchanging the two brane tensions,
\(\Sigma_1\longleftrightarrow\Sigma_2\), corresponds to
\(\alpha_L\longleftrightarrow\alpha_R\). The crossing entanglement
entropy is invariant under this exchange even when \(l_L\) and
\(l_R\) are held fixed. This is not a spatial reflection, since
such a reflection would also exchange the interval endpoints.
The invariance can be seen directly by eliminating the auxiliary
geometric quantities from \eqref{eq:rel2brane}. To see this, define
$\kappa\equiv\sin\varphi_{CR}/\sin\alpha_R
=
\sin\varphi_{CL}/\sin\alpha_L$. The exterior-circle relations then give
\begin{equation}
D_R=\frac{l_R}{1+\kappa},
\qquad
D_L=\frac{l_L}{1-\kappa}.
\end{equation}
Equating the two expressions for \(D_C\) gives
\(\kappa\) :
\begin{equation}
\frac{l_R}{l_L}
=
\frac{1+\kappa}{1-\kappa}\cdot
\frac{\sin(\alpha_R-\varphi_{CR})}
     {\sin(\alpha_R+\varphi_{CR})}\cdot
\frac{\sin(\alpha_L-\varphi_{CL})}
     {\sin(\alpha_L+\varphi_{CL})}.
\end{equation}
Expressing $\varphi_{CL}$ and $\varphi_{CR}$ in terms of $\kappa$ on the branch where 
\(\cos\varphi_{CR},\cos\varphi_{CL}\geq 0\) yields
$\varphi_{CR}=\arcsin\!\bigl(\kappa\sin\alpha_R\bigr)$,
$\varphi_{CL}=\arcsin\!\bigl(\kappa\sin\alpha_L\bigr)
$. Substituting into the endpoint ratio gives
\begin{equation}
\frac{l_R}{l_L}
={}
\frac{1+\kappa}{1-\kappa}\cdot
\frac{
\sqrt{1-\kappa^2\sin^2\alpha_R}-\kappa\cos\alpha_R
}{
\sqrt{1-\kappa^2\sin^2\alpha_R}+\kappa\cos\alpha_R
}
\times
\frac{
\sqrt{1-\kappa^2\sin^2\alpha_L}-\kappa\cos\alpha_L
}{
\sqrt{1-\kappa^2\sin^2\alpha_L}+\kappa\cos\alpha_L
}.
\end{equation}
This equation is manifestly invariant under
\(\alpha_R\longleftrightarrow\alpha_L\), without exchanging \(l_R\) and
\(l_L\). The same value of \(\kappa\) therefore solves the equations after
the exchange, while the angular solutions are mapped according to
\(\varphi_{CR}\longleftrightarrow\varphi_{CL}\).
Using \eqref{eq:gammar}-\eqref{eq:gammac}, the entropy may similarly be written as 
\begin{equation}
S_\text{A}
={}
\frac{R}{4G}
\log\Biggl[
\frac{4l_Rl_L}{\epsilon^2(1-\kappa^2)}\cdot
\frac{
\sqrt{1-\kappa^2\sin^2\alpha_R}-\cos\alpha_R
}{
\sqrt{1-\kappa^2\sin^2\alpha_R}+\cos\alpha_R
}
\times
\frac{
\sqrt{1-\kappa^2\sin^2\alpha_L}-\cos\alpha_L
}{
\sqrt{1-\kappa^2\sin^2\alpha_L}+\cos\alpha_L
}
\Biggr],
\end{equation}
where we have used the identity $\tan \frac{x+y}{2} \tan\frac{x-y}{2} = \frac{\cos y - \cos x}{\cos y+\cos x}$.
The dependence on the two tensions thus occurs through a symmetric product,
which explains why exchanging \(\alpha_R\) and \(\alpha_L\) leaves the
entropy invariant although the interval endpoints are not exchanged. 
Using \(R\Sigma_1=-2\cos\alpha_L\) and
\(R\Sigma_2=-2\cos\alpha_R\), this establishes the invariance under
exchange of the brane tensions noted after
\eqref{eq::twobranesrhocs}.

%%%%%%%%%%%%%%%%%%%%%%%%%%%%

\addcontentsline{toc}{section}{References}

\bibliography{biblio.bib}
\bibliographystyle{JHEP}

\end{document}